\documentclass[9pt,twocolumn,twoside]{osajnl}

\journal{josaa} 
\usepackage{widetext}
\setboolean{shortarticle}{false} 

\ifthenelse{\boolean{shortarticle}}{\colorlet{color2}{color2b}}{\colorlet{color2}{color2}} 

\title{Crossing statistics of scattered Laser Light through Nanofluid}

\author[1,2]{M. Arshadi Pirlar}
\author[1,*]{S.M.S. Movahed}
\author[2]{D. Razzaghi}
\author[1]{R. Karimzadeh}

\affil[1]{Department of Physics, Shahid Beheshti University, Velenjak, Tehran 19839, IRAN}
\affil[2]{Photonic and Quantum Technology Research School, NSTRI, 11155-3486, Tehran, IRAN}

\affil[*]{Corresponding author: m.s.movahed@ipm.ir}

\dates{Compiled \today}

\ociscodes{(000.5490) Probability theory, stochastic processes, and statistics; (120.5820) Scattering measurements; (030.6140) Speckle; (030.6600) Statistical optics; (120.7250) Velocimetry; (030.5770) Roughness.}

\doi{\url{http://dx.doi.org/10.1364/ao.XX.XXXXXX}}

\begin{abstract}
In this paper, we investigate the crossing statistics of speckle patterns formed in Fresnel diffraction region by a laser beam scattering through a nanofluid. We extend $zero-crossing$ statistics to assess dynamical properties of nanofluid. According to joint probability density function of laser beam fluctuation and its time derivative, theoretical framework for Gaussian and non-Gaussian regimes are revisited. We count number of crossings not only at {\it zero} level but also for all available thresholds to determine the average speed of moving particles. Using probabilistic framework in determining crossing statistics, {\it a priori} Gaussianity is not essentially considered, therefore even in presence of deviation from Gaussian fluctuation, this modified approach is capable to compute relevant quantities such as mean value of speed more precisely. Generalized total crossing which represents the weighted summation of crossings for all thresholds to quantify small deviation from Gaussian statistics is introduced. This criterion can also manipulate the contribution of noises and trends to infer reliable physical quantities. The characteristic time scale for having successive crossings at a given threshold is defined. In our experimental setup, we find that increasing sample temperature leads to more consistency between Gaussian and perturbative non-Gaussian predictions.  The maximum number of crossing does not necessarily occur at mean level indicating that we should take into account other levels in addition to {\it zero} level to achieve more accurate assessments.
\end{abstract}

\setboolean{displaycopyright}{true}

\begin{document}

\maketitle
\thispagestyle{fancy}

\ifthenelse{\boolean{shortarticle}}{\ifthenelse{\boolean{singlecolumn}}{\abscontentformatted}{\abscontent}}{}

\section{Introduction}

Complexities are ubiquitous in various phenomena due to initial conditions and evolutions.  Quantifying relevant observable quantities essentially depends on exploiting robust statistical approaches. Probabilistic framework corresponding to a typical statistical measure, includes joint probability density function (JPDF) of relevant dependent parameters and we should determine mentioned JPDF for further statistical inference. For a typical stochastic field, $x$, in $(1+D)$-Dimension (the index "$1$" refers to $x$ and "$D$" represents the number of independent parameters), statistical properties is fully characterized by JPDF as $\mathcal{P}(\bf A)$ and ${\bf A}\equiv\{x,\partial_ix,\partial_{ij}x,...\}$ \cite{adler81,adler07}.

Among different statistical quantities to characterize the morphology of a stochastic fluctuation in various dimensions, a promising class is devoted to crossing statistics. This
method can be used for 1, 2 and 3 dimensional fluctuations
corresponding to crossing statistics, length or contour statistics and area statistics, respectively. Mentioned measures have been extensively utilized and improved after pioneering study done by  S. O. Rice in the context of mathematical analysis of random noise  \cite{rice44}. Comprehensive explanation to evaluate geometrical and topological nature of stochastic fields has been given in \cite{adler81,adler07}. 
 This notion has also opened new trends in examining stochastic fields ranging from condensed matter and surface physics, optics to astronomy and cosmological random fields. P. H. Brill et al, provided a brief tutorial on how to apply this method in various stochastic models, such as queues, inventories, dams, risk reserve models in insurance, counter models \cite{percy00}. We can notice to application of level crossing analysis for the returns of market prices \cite{jafari06} and a wide range of fractional Gaussian noises \cite{newland,Vahabi}. Statistical properties of many radiophysical problems  have been investigate in \cite{Khimenko77}. The surface growing processes were examined in \cite{shahbazi03}. Also the fluctuations of velocity field in the Burgers turbulence were measured \cite{burger06} and the characterization of anisotropy in a 2D field, e.g. surface height fluctuations based on crossing statistics has been done in \cite{Ghasemi15}. The crossing statistics of the amplitude and the intensity of a monochromatic speckle pattern as well as the behavior of the spatial derivatives of these quantities have been studied theoretically in \cite{ebel79,barakat80,bahu80}, and a theoretical limits on the light distributions with random phase diffusers were derived in \cite{kurtz73,Goodman06}. Laser speckle velocimetry based on the {\it zero-crossing} rate of the spatially integrated speckle intensity variations under various optical configurations of illumination and detection has been studied in \cite{Asakura,Takai,Iwai,Asakura81,Takai2}. R. Barakat presented a closed form for the {\it zero} and level-crossing rate of differentiated speckle \cite{Barakat,Barakat88}. Application of level crossing analysis was developed for various probability density functions with primary relevance for optics \cite{Yura10}. It was shown that the sought-for probability density can be expressed in terms of modified Bessel functions of the second kind \cite{Barakat91}. 
Accordingly, crossing statistics has been used  to quantify area of isodensity contours of cosmological models and large scale structure \cite{Ryden89,ryden1988}. Application of crossing statistics for redshift space and for other complementary perturbative models have been given in   \cite{Matsubara96,matsubara03}. Also to detect cosmic strings networks as a type of topological defects, level crossing has been utilized in \cite{sadegh11}.

 A motivation behind crossing statistics is devoted
to possibility derivation of an explicit theoretical prediction
based on JPDF of random variables and
corresponding derivative with respect to a dynamical variable
\cite{Beckmann67}. A main goal to utilize crossing statistics
in various field of researches is examining the capability of number
of crossing at a typical threshold to quantify underlying fluctuations. As explained in details  by Yura et al., in optical communication and sensors systems, the statistical properties of fade and surge play key role in characterizing associated physical behavior \cite{Yura10}.

In recent years, extensive researches have been conducted on applications of nano-materials in nanofluids technologies. Nanofluid is a new kind of heat transfer fluid prepared by dispersing nanoparticles in traditional based fluids \cite {Kakac1, Kakac}. It is shown that they can enhance effective heat transfer properties of the original base fluid \cite {Wong, Duang}. The most common nanoparticles for nanofluids are generally metallic and nonmetallic materials such as Al$_2$O$_3$, Ag, CuO, Cu, SiO$_2$ and TiO$_2$  \cite {Azmi, Buongiorno, Wang}. The movement of nanoparticles plays a significant role in the anomalously increased heat transfer properties of the nanofluids \cite{Guyot,Chicea}. So, it is of great importance to find out a robust method to measure the movements of nanoparticles in nanofluids, in order to understanding the heat transfer enhancement mechanism.

Due to the complexity of nanoparticles movement in nanofluids, calculating the velocity of each particles in the fluid is impossible, but using statistical methods, the average speed of particles can be computed with proper experimental configuration \cite {Asakura,Asakura81}. When a coherent laser light passes through a fluid, a non-uniformly illuminated image (speckle pattern) can be formed. The speckle pattern appearances due to the interference of the scattered light with different phases and amplitudes by the nanoparticles. This pattern changes in time as a consequence of the nanoparticle motion due to the aggregation, sedimentation and Brownian motion \cite{ Guyot, Chicea}. Therefore, the speckle pattern analysis can be used to characterize the dynamical behavior of nanofluids. 

Some previous studies have almost focused on fluctuations of light intensity around mean level which is so-called {\it zero-crossing} \cite{Asakura,Takai, Iwai,Asakura81,Takai2,Barakat}. In mentioned method, the spatially integrated speckle intensity fluctuations produced by moving diffuse particles in a plane is investigated by counting the crossing at mean level. In addition,  direction of moving particles can be determined by using directional detecting aperture \cite{Iwai}. Central limit theorem is a key  assumption for deriving theoretical {\it zero-crossing} function leading to an expression including the moving particle's speed in simple form and now becomes almost useful real time method. Generally, deviation from Gaussianity is occurred in a typical statistical fluctuations, therefore, we should take into account non-Gaussian contributions. Another alternative method to reveal mean-speed of suspended particle in a fluid is Spatial Filtering Velocimetry \cite{aizu}. In this approach, Fourier space analysis is utilized to compute speed of diffuse particles using optical approach and consequently experimental setup to achieve accurate measurement  is somehow difficult \cite{Asakura81,aizu}. 

In this paper, we construct an experimental setup to create 
speckle patterns. Such patterns are formed in Fresnel diffraction region by illumination of a laser beam on the nanoparticles suspended in a nanofluid. As a movement of the nanoparticles, the speckle pattern is not static and varies with time. In this experimental setup, we use a He-Ne laser light at 633 nm wavelength as a light source. 
By measuring the spatially integrated intensity fluctuation of the speckles and its variations with time, we are able to examine the movement characteristics of the nanofluids \cite{Asakura,Takai, Iwai, Asakura81,Takai2,Barakat}.

Relying on previous statistical analysis of speckle pattern due to scattering laser light from moving diffuse particles used to determine the magnitude and direction of moving objects \cite{Asakura,Takai, Takai2, Iwai}, present research has following advantages and novelties:  we try to use generalized version of {\it zero-crossing} which is so-called crossing statistics enumerating crossing for all available thresholds to achieve more precise measurement of relevant quantities, particularly, mean value of speed of diffuse particles. Non-Gaussianity due to the various physical features of moving diffuse particles as well as experimental setup can now be taken into account, theoretically, to find feasible results. Perturbation approach in the context of Edgeworth expansion will be utilized to compute the mean-speed of diffuse particles in weak non-Gaussian case. It is worth noting that for optical applications, previous efforts have been concentrated on deriving JPDF for relevant fluctuations in closed form resulting in analytical expression for level-crossing statistics. While in the present paper, we exploit the perturbative expansion of JPDF known as Edgeworth expansion \cite{Scherrer91,Juszkiewicz95,Bernardeau95,matsubara03}. Therefore, this approach enables us to investigate various processes irrespective to existence of necessary knowledge concerning functional form of JPDF.  In addition we introduce generalized total crossing which is a suitable benchmark for assessment of large and small fluctuations, separately. Finally we will also introduce characteristic time scales in such experimental setup referring to crossing statistics framework. 


The rest of this paper is organized as follows: In Section \ref{model}, we will explain theoretical background of crossing statistics and its generalizations. Gaussian and non-Gaussian fluctuations will be considered in the context of crossing statistics. Data description and experimental setup to collect data sets will be revealed in section \ref{data}.  Section \ref{application} is devoted to applications of crossing statistics on intensity fluctuation of scattered laser light through a nano-fluid. In section \ref{result}, we will give results and relevant results achieved in this paper.  Concluding remarks will be given in section \ref{summary}.


\section{Theoretical Model: Up- and down-crossing statistics}\label{model}

S. O. Rice introduced level crossing or generally crossing statistics \cite{rice44}. Soon after that,  in various disciplines raging from complex systems  \cite{percy00,newland,Khimenko77,jafari06,Vahabi}, material sciences \cite{Ghasemi15,shahbazi03,burger06}  optics \cite{ebel79,barakat80,bahu80,kurtz73,Goodman06,Asakura,Takai,Iwai,Asakura81,Takai2,Barakat,Barakat88,Yura10,Barakat91} to cosmology and early universe  \cite{Ryden89,ryden1988,Matsubara96,sadegh11,matsubara03},  mentioned  method has been used and improved. 
Crossing statistics represents geometrical properties of a typical stochastic process, therefore, it has proper capability in order to quantify fluctuations in a robust manner.

\subsection{Mathematical framework}\label{sec:math}

\begin{figure}[htbp]
\centering
\fbox{\includegraphics[width=\linewidth]{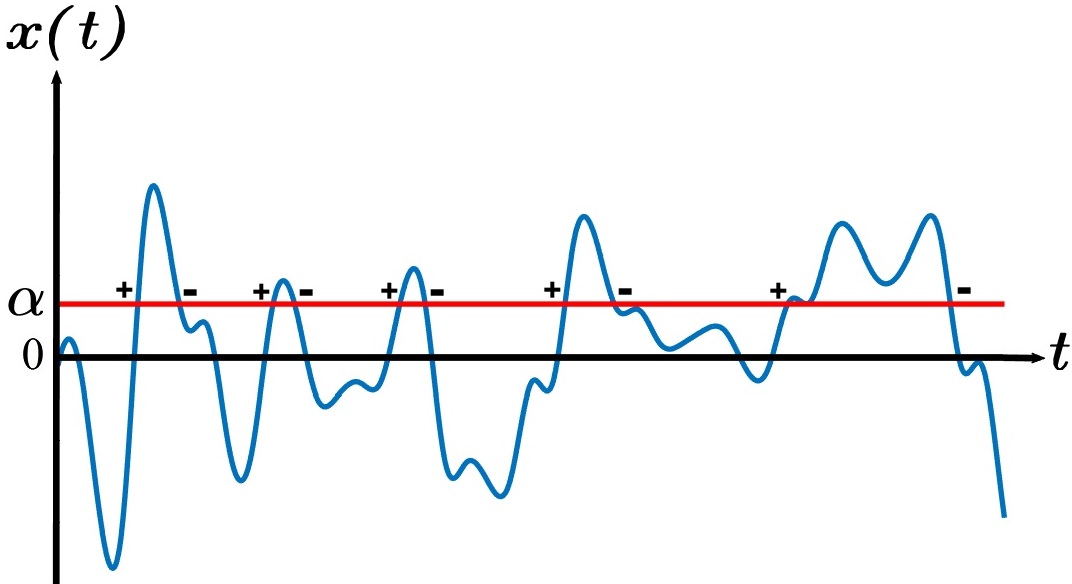}}
\caption{A sketch for positive and negative slope crossings of a typical time series at a given threshold $\vartheta\equiv\alpha/\sigma_0$.}
\label{fig:1}
\end{figure}

A discrete set of typical fluctuations in $(1+1)$-Dimension recorded in an experiment is represented by: $\{x(t_i)\},\quad i=1,..,N$, and $\Delta t\equiv
t_{i+1}-t_{i}$. For convenient, we set the mean value of recorded series to
zero, $\langle x(t)\rangle=0 $. In continuous limit, up- and down-crossing are defined according to
crossing points with positive and negative slopes at an arbitrary
threshold, $\vartheta\equiv\alpha/\sigma_0$, respectively. Where 
$\sigma_0^2=\langle x(t)^2\rangle$ and 
$\alpha$ is a typical value in domain of $\{x(t_i)\}$ (see 
Fig. \ref{fig:1} for more details). 
The crossing number density of time series at a given threshold, $\vartheta$, can be written as $n_{x}^{\pm}(t;\alpha)$ (see Fig. \ref{fig:1} for more details). We consider "$+$" for up-crossing and "$-$" for down-crossing. Mathematical description of ensemble average of crossing point is:
\begin{eqnarray}
\label{eqn1}
\langle n_{x}^{\pm}(t;\alpha)\rangle &=&\sum_{t_{\pm}}\delta_D(t-t_{\pm})
\end{eqnarray}
here $\delta_D$ is Dirac delta function. 
An important task to do is making a connection between Dirac delta function and corresponding variables in Eq. (\ref{eqn1}). To this end, we can expand $x(t)$ around a typical time associated with up- or down-crossings as follows:
\begin{equation}
x(t)=x(t_{\pm})+(t-t_{\pm})\ \eta(t_{\pm})+\mathcal{O}(\delta t^2)
\end{equation}
 here $\eta\equiv\partial_t x$. For crossing at $\alpha$, we should set $x(t_{\pm})=\alpha$, therefore, one can find a useful expression for $\delta_D(t-t_{\pm})=\delta_D(x(t)-\alpha)|\eta(t_{\pm})|$. Inserting mentioned expression in Eq. (\ref{eqn1}) results in:
\begin{eqnarray}
\label{eqn2}
\langle n_{x}^{\pm}(t;\alpha)\rangle &=&\int dx \int d \eta\ \delta_D(x(t)-\alpha)\ |\eta(t_{\pm})|\ {\mathcal{P}}(x,\eta)
\end{eqnarray}
To have up-crossing (crossing with positive slope) additional
constraint should be taken into account on first derivative of
record data as $\Theta(\eta)$ ($\Theta(:)$ is step function), while
for down-crossing (crossing with negative slope) we have
$\Theta(-\eta)$. Using the functional form of JPDF, theoretical
prediction for up- and down-crossings are achieved. 
Finally, number density of crossing has the following form:
\begin{eqnarray}\label{theory for nu gaussain}
\langle n_{x}^{\pm}(\alpha)\rangle&=&\left \langle \delta_D(x(t)-\alpha)\Theta (\pm\eta)|\eta| \right \rangle
\end{eqnarray}
 The total expected number of crossings of underlying time series per unit time at the threshold $\vartheta=\alpha/\sigma_0$ for all possible values of $\eta$ reads as:
 \begin{eqnarray}
\label{eq9}
N_{\alpha}&=&\int_{-\infty}^{\infty} d\eta|\eta| \mathcal{P}(\alpha,\eta)\\
&=&\langle n_{x}^{+}(\alpha)\rangle+\langle n_{x}^{-}(\alpha)\rangle
\nonumber
\end{eqnarray}
Above equation  represents the expected statistical crossings at threshold $\vartheta=\alpha/\sigma_0$ with both positive and negative slopes as shown in Fig. \ref{fig:1}. Number of crossing at  {\it zero} level  which is so-called {\it zero-crossing} for Gaussian process is given by:
\begin{eqnarray}
\label{eq10}
N_{0}&=&\langle n_{x}^{+}(0)\rangle+\langle n_{x}^{-}(0)\rangle=\mathcal{P}(0)\ \langle|\eta|\Theta(+\eta)+|\eta|\Theta(-\eta)\rangle\nonumber\\
\end{eqnarray}
For a stationary Gaussian statistics, we have: $\mathcal{P}(x,\eta)=\mathcal{P}(x)\mathcal{P}(\eta)$, therefore by using Eqs. (\ref{eq9}) and (\ref{eq10}) we have:
 \begin{eqnarray}
\label{eq11}
N_{\alpha}=N_{0}\ \frac{\mathcal{P}(\alpha)}{\mathcal{P}(0)}
\end{eqnarray}
For the generalized total level crossing, $N_{total}$, we have \cite{jafari06,Vahabi}:
\begin{eqnarray}
\label{eq37}
N_{total}(q)=\int_{-\infty}^{\infty} N_{\alpha}|\alpha-\langle x(t)\rangle |^qd\alpha 
\end{eqnarray}
where $q$ represents the order of generalized moment. It turns out, for $q=0$, the quantity $N_{total}(0)$ equates to total number of crossing for all thresholds. Therefore $N_{total}(0)$ reveals  a measure for roughness of underlying signal. Eq. (\ref{eq37}) represents the general definition of crossing statistics. It is possible to examine the contribution of various size of fluctuations according to $N_{total}(q)$ quantity.  For $q<1$,  small fluctuations have dominant impact, on the contrary, for $q\ge1$, large fluctuations play crucial role in computing generalized crossing statistics. Subsequently, weighted analysis is systematically retrieved by this quantity. This measure is also very sensitive to any deviation form Gaussian statistics. In addition it is useful to manipulate  the contribution of noises and trends in a recorded series.
 
 In what follows, we are going to examine Gaussian and non-Gaussian processes.
\begin{figure}[t]
\centering
\includegraphics[width=\linewidth]{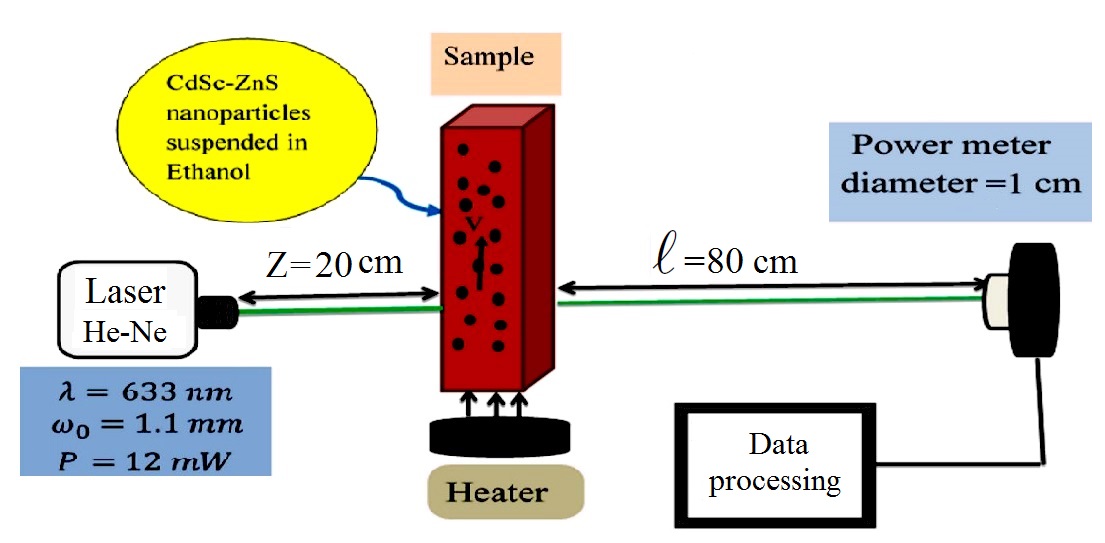}\\
\includegraphics[width=9cm, height=5cm]{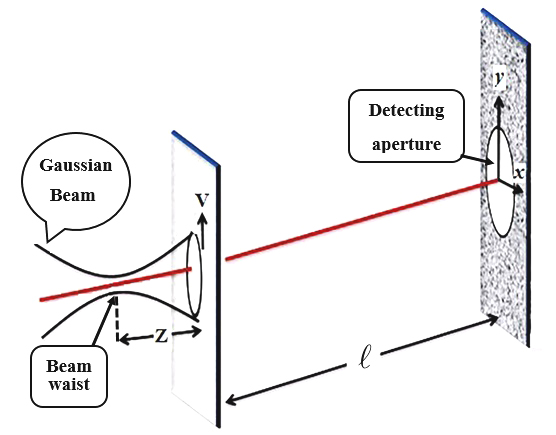}
\caption{Upper panel corresponds to schematic diagram of a combined system consisting of the optical arrangement of the spatially integrated speckle intensity. In lower panel we show a sketch of the optical arrangement for the formation of speckles in the diffraction field due to a moving diffuse nanoparticle under illumination of a Gaussian beam. In this figure, the circular aperture is shown to detect spatially integrated speckles.}
\label{fig:3}
\end{figure}
\subsection{Gaussian fluctuation}
In previous subsection, we derived  integration form for crossing statistics. In this subsection, we assume a Gaussian process with variables ${\bf A}\equiv\{ x, \eta\}$, subsequently, the functional form of JPDF can be considered as bivariate Gaussian function:
\begin{eqnarray}\label{JPDF11}
{\mathcal P}_G({\bf{A}})=\frac{1}{2\pi\sqrt{Det({\mathcal{M}})}}
{\bf e}^{-\frac{1}{2}({\bf A}^{T}.\mathcal{M}^{-1}.{\bf A})}
\end{eqnarray}
where $\mathcal{M}$ is the covariance matrix of underlying variables:
\begin{eqnarray}\label{cov1}
\mathcal{M}\equiv {\rm Cov}= \left[\begin{array}{cc}
\langle x^2\rangle & \langle x\eta\rangle  \\
\langle\eta x\rangle & \langle\eta^2\rangle  \end{array} \right].
\end{eqnarray}
Each element of covariance matrix can be computed using the power
spectrum, $S(\omega)$, of given process,  accordingly:
\begin{eqnarray}
C_{xx}(\tau)\equiv\langle x(t)x(t+\tau)\rangle&=&\frac{1}{2\pi}\int d\omega\ {\bf e}^{i\omega \tau}S(\omega)\\
 C_{xx}(0)\equiv\sigma_0^2=\langle x^2\rangle&=&\frac{1}{2\pi}\int d\omega S(\omega)\\
C_{\eta\eta}(0)\equiv \sigma_1^2=\langle \eta^2\rangle& =&\frac{1}{2\pi}\int d\omega \omega^2 S(\omega)
\end{eqnarray}
where $\sigma_0$ and $\sigma_1$ are called spectral indices. In addition $C_{\eta\eta}=-C_{x\xi}=-\langle x(t)\xi(t+\tau)\rangle$, here $\xi(t)=\partial^2 x(t)/\partial t^2$. For stationary time series $\langle x\eta\rangle=0$, 
consequently, Eq. (\ref{theory for nu gaussain}) becomes:
\begin{eqnarray}\label{eq:n3}
\langle n_{x}^{\pm}(\alpha)\rangle &=& \frac{1}{2\pi}\frac{\sigma_1}{\sigma_0}\ {\bf e}^{-\alpha^2/2\sigma_0^2}
\end{eqnarray}\
therefore Eqs. (\ref{eq10}) and (\ref{eq11}) become:
\begin{eqnarray}
\label{eq23}
N_{0}&=&\frac{1}{\pi}\frac{\sigma_1}{\sigma_0}
\end{eqnarray}
\begin{eqnarray}
\label{eq233}
N_{\alpha}&=&N_{0}\ {\bf e}^{-\alpha^2/2\sigma_0^2}
\end{eqnarray}
Plug in $\sigma_{0}$ and $\sigma_{1}$ into Eq. (\ref{eq23}), we have:
\begin{equation}
\label{eq26}
N_{0}=\frac{1}{\pi} \left(-\frac{C_{x\xi}(0)}{C_{xx}(0)}\right)^{\frac{1}{2}}
\end{equation}
This relation shows that the number of {\it zero-crossing} per unit time is proportional to the square root of second derivative autocorrelation function divided by autocorrelation function.  For number of crossing at a given threshold one can write:
\begin{equation}
\label{eq27}
N_{\alpha}=\frac{1}{\pi} \left(-\frac{C_{x\xi}(0)}{C_{xx}(0)}\right)^{\frac{1}{2}}{\bf e}^{-\alpha^2/2C_{xx}(0)}
\end{equation}
In addition the generalized total crossing introduced by Eq. (\ref{eq37}) for Gaussian process is:
\begin{eqnarray}
\label{eq277}
N_{total}(q)&=& \frac{(2C_{xx}(0))^{\frac{q+1}{2}}}{\pi} \sqrt{-\frac{{C_{x\xi}(0)}}{{C_{xx}(0)}}}\Gamma \left(\frac{q+1}{2}\right)
\end{eqnarray}

Following subsection is devoted to non-Gaussian derivation for crossing statistics.

\subsection{Non-Gaussian fluctuations: Perturbation approach}
In pervious subsection, we derived the analytical function for $\langle n_{x}^{\pm}(\alpha)\rangle$ for a perfect Gaussian process. There are many reasons causing to have non-Gaussianity, therefore in context of perturbation approach, T. Matsubara tried to calculate an expansion for crossing statistics in terms of various order of $\sigma_0$ \cite{matsubara03}. To this end, we start from characteristic function which is Fourier transform of JPDF as:
\begin{eqnarray}
\mathcal{Z}_{\bf A}(\Lambda)=\int d{\bf A} \mathcal{P}(\bf A){\bf e}^{-i\Lambda \cdot {\bf A}}
\end{eqnarray}
Using the cumulant expansion theorem \cite{ma85}, above partition function can be read as: 
\begin{eqnarray}
\mathcal{Z}_{\bf A}(\Lambda)={\bf e}^{-\frac{1}{2}\Lambda ^T\cdot {\mathcal{M}}^{-1}\cdot \Lambda}{\bf e}^{\sum_{n=3}^{\infty}\frac{i^n}{n!}  \sum_{\mu_1...\mu_{n}}\mathcal{M}^{(n)}_{\mu_1...\mu_n} \Lambda_{\mu_1}...\Lambda_{\mu_n}}
\end{eqnarray}
here $\mathcal{M}^{(n)}_{\mu_1...\mu_n}$ is cumulant. Inverse Fourier transform of partition function leads to:
\begin{eqnarray}
\mathcal{P}({\bf A})={\bf e}^{\sum_{n=3}^{\infty}\frac{(-1)^n}{n!}  \sum_{\mu_1...\mu_{n}}\mathcal{M}^{(n)}_{\mu_1...\mu_n} \frac{\partial^n}{\partial A_{\mu_1}...\partial A_{\mu_n}}}\mathcal{P}_G({\bf A})
\end{eqnarray}
and $\mathcal{P}_G({\bf A})$ is multivariate Gaussian function (see Eq. (\ref{JPDF11})). 
 The mean value of a given statistical quantity, $\mathcal{F}(\bf A)$, can be expressed by ensemble average as: 
\begin{eqnarray}\label{fper}
\langle \mathcal{F}\rangle&=&\int d{\bf A} \mathcal{P}({\bf A})\mathcal{F}(\bf A)\nonumber\\
&=&\left \langle  {\bf e}^{\sum_{n=3}^{\infty}\frac{(-1)^n}{n!}  \sum_{\mu_1...\mu_{n}}\mathcal{M}^{(n)}_{\mu_1...\mu_n} \frac{\partial^n}{\partial A_{\mu_1}...\partial A_{\mu_n}}} \mathcal{F}(\bf A)  \right \rangle_{G}
\end{eqnarray}
in which $\langle \rangle_G\equiv \int d{\bf A}\mathcal{P}_G(\bf A)\mathcal{F}(\bf A)$. As an illustration, probability density function for weak non-Gaussianity which is known as Edgeworth expansion becomes \cite{matsubara03}:
\begin{eqnarray}\label{pper}
\mathcal{P}(\alpha)&=&\langle \delta_{D}(x-\alpha)\rangle\nonumber\\
&=& \frac{{\bf e}^{-\alpha^2/2\sigma_0^2}}{\sqrt{2\pi}\sigma_0}\left[1+\sigma_0\frac{S^{(0)}}{6}H_3(\alpha/\sigma_0)+\mathcal{O}(\sigma_0^2)\right]
\end{eqnarray}
here $S^{(0)}\equiv \langle x(t)^3\rangle/\sigma_0^4$ and $H_3$ is probabilistic Hermite polynomial of rank 3. The Hertmite polynomials (well-known in probability theory) as a complete orthogonal set have been used  to expand PDF and the corresponding coefficients of expansion are cumulants which reveal the statistical properties of underlying process. Finally, we obtain a proper expansion for a non-Gaussian distribution in terms of corresponding cumulants.  According to Eqs. (\ref{fper}) and (\ref{pper}), the perturbative functions for crossing statistics \cite{matsubara03} and generalized total crossing are derived as:
\begin{widetext}
\begin{eqnarray}
\label{pertur1}
N_{\alpha}&=&\frac{1}{\pi} \sqrt{-\frac{C_{x\xi}(0)}{C_{xx}(0)}}{\bf e}^{-\alpha^2/2C_{xx}(0)}
\left\{1+\sigma_0\left[\frac{S^{(0)}}{6}H_3(\alpha/\sigma_0)+\frac{S^{(1)}}{3}H_1(\alpha/\sigma_0)\right]+\mathcal{O}(\sigma_0^2)\right\}\\
N_{total}(q)&=& \frac{2^{(q+1)/2}\sigma_1\sigma_0^q}{\pi}\left\{  \Gamma\left(\frac{q+1}{2}\right)+ \sigma_0\left[\frac{(2+q)\sigma_0^2-3}{6}S^{(0)}+\frac{S^{(1)}}{3}\right]\frac{q}{2}\Gamma\left(\frac{q}{2}\right)+\mathcal{O}(\sigma_0^2)\right\}
\end{eqnarray}
\end{widetext}
where $S^{(1)}\equiv -\frac{3}{4}\langle x(t)^2\xi(t)\rangle/(\sigma_0^2\sigma_1^2)$ and $\Gamma(:)$ is gamma function.
In this section we derived general definition of crossing statistics for a given threshold, subsequently, an extension for computing relevant physical quantities is achieved. In order to use perturbative expansion, second cumulant should be remained finite. In section  \ref{application}, we will apply this approach on experimental data sets and compute mean-speed of diffuse objects.  We will also examine the Gaussianity nature of recorded fluctuations in our experimental setup.

Inspired by formulation for
fully-developed turbulent flows, another interesting
method to quantify non-Gaussian PDF has been introduced. This approach was originally
illustrated to examine fully developed turbulence
\cite{cas90,chab94,aren98,bac01} and recently has been used in wide
range of researches such as foreign exchange rate
\cite{Ghashghaie96}, stock index \cite{kiy06}, and human heartbeat
\cite{kiy04,kiy05}, sol-gel transition \cite{sol10,sol13} and
petrophysical quantities \cite{kohi15}. It is also useful to apply this method to examine non-Gaussian behavior of underlying process.

\section{Experimental setup and data description} \label{data}
Our versatile experimental setup to examine the intensity of scattered light through nanofluid is conducted by using a combined systems represented in upper panel of Fig. \ref{fig:3}. It consists of the optical arrangement for the formation of speckles and the electrical processing of the spatially integrated speckle intensity. The Gaussian beam generated  by a single mode He-Ne laser with the wavelength $\lambda=633 nm$ with beam waist $ w_0=1.1 mm$ is focused on the sample. Our sample is located with a distance $z=20 cm$ away from the position of the beam waist. The speckle patterns are formed by illuminating the CdSc-ZnS nanoparticles suspended in ethanol. These patterns are detected in the Fresnel diffraction field at the plane, a distance $\ell=80 cm$ away from the center of sample by a circular aperture with a radius $d=5 mm$ in front of power meter. To achieve  a drift speed due to convection, a heater is located under the sample. (Fig. \ref{fig:3}).

\begin{figure*}[ht]
\begin{center}
\includegraphics[width=4.5cm, height=4cm]{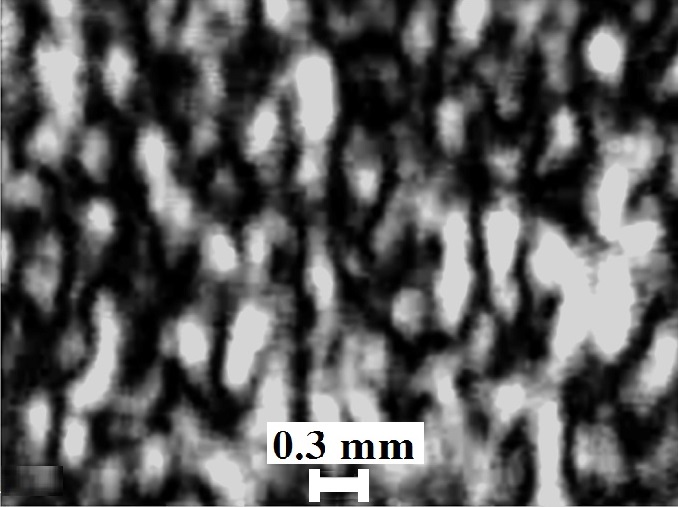}
\includegraphics[width=4.5cm, height=4cm]{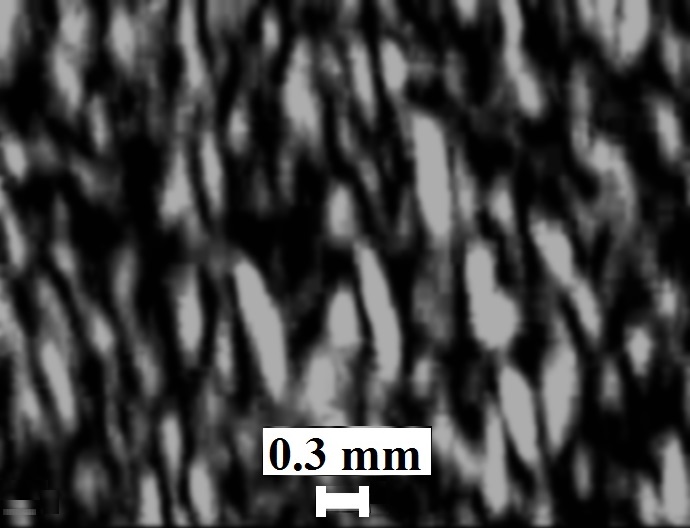}
\includegraphics[width=4.5cm, height=4cm]{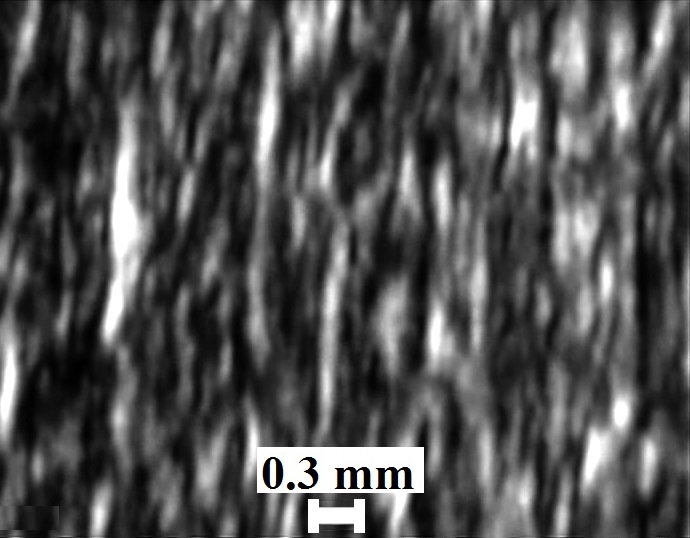}
\includegraphics[width=14cm, height=6cm]{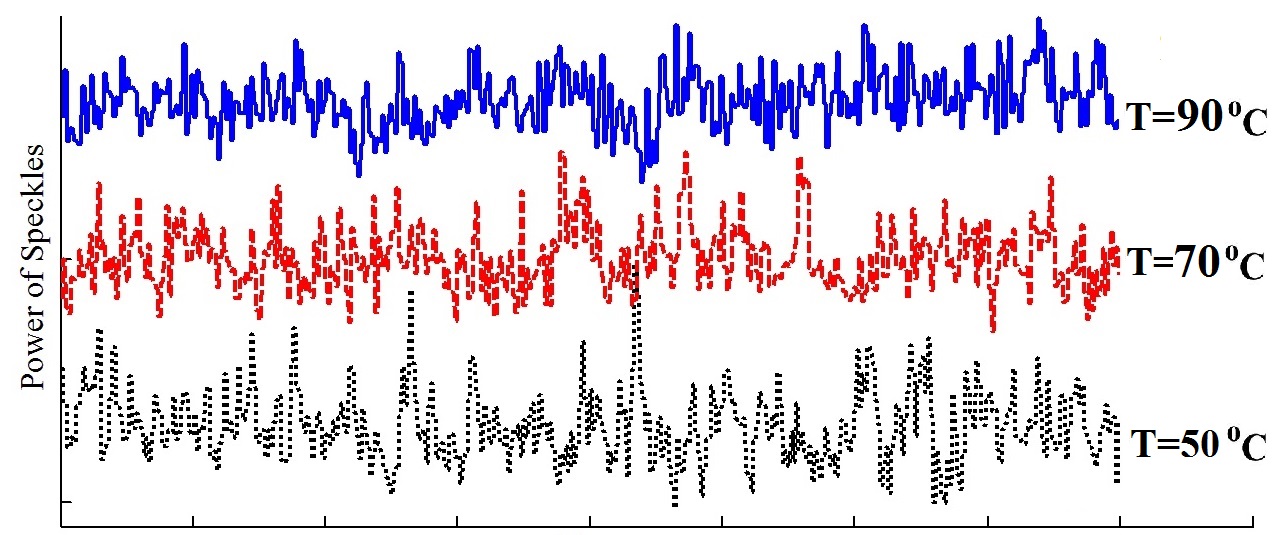}
\caption{\label{fig:4}Upper panel indicates speckles patterns recorded by CCD camera from left to right corresponds to $T=50^{\circ}$C, $T=70^{\circ}$C and $T=90^{\circ}$C, respectively. Lower panel corresponds to power of integrated speckle versus time for three different experiments.}
\end{center}
\end{figure*}

If a coherent light is incident upon a nanofluids containing
randomly distributed nanoparticles, a nonuniform illuminated image
is obtained. This pattern is generated due to the interference of
the scattered lights by the nanoparticles. Subsequently,
fluctuations of the image intensity in each location of the
interference field, which is so-called   "{\it speckle}" is
appeared.
 As a movement of the nanoparticles, the speckle pattern is not static and changes with time. So one can consider the time-varying of speckle intensity fluctuation is a stochastic field.
 Power meter records integrated intensity of all speckles during a given interval time which is called spatially integrated speckle intensity. We define the fluctuations of spatially integrated speckle intensity by $I(t)-\langle I(t)\rangle$ where $\langle\rangle$ denotes the ensemble average of the speckle intensity variation. In the lower panel of Fig. \ref{fig:3} we indicate the optical system for the formation of speckles under the illumination of a Gaussian laser beam over a transmitting diffuse object moving with constant speed in a certain direction of a plane perpendicular to the optical axis.
 Upper panel of Fig. \ref{fig:4} illustrates  the formation of speckles intensity recorded by CCD camera. Lower panel of Fig. \ref{fig:4} represents power fluctuations as a function of time for three different experiments with different physical conditions. These data sets will be used as input for crossing statistics explained in previous section in order to determine magnitude of diffuse object's speed and relevant statistical properties of intensity fluctuation containing some information about the dynamics of speckle. To measure the direction of diffuse object's speed, we must utilize rotating directional detecting aperture \cite{ Iwai}.

\section{Implementation on scattered Laser Beam}\label{application}
One of the main purpose of this analysis is devoted to determining the average of diffuse speed of nanoparticle suspended in a solution. The {\it zero-crossing} analysis of the dynamic speckles can be used to measure the speed of nanoparticles in nanofluids. According to this method, the crossing statistics of intensity fluctuations of speckles at {\it zero} level corresponding to mean value, $\langle I(t)\rangle$ is enumerated \cite{Asakura,Asakura81}.  To get reliable result, two following conditions should be satisfied. Firstly, the convection speed of nanofluid must be constant and perpendicular to the direct of Gaussian laser beam. To achieve such situation, a heater is placed at a position in beneath of sample. Therefore, one can control the temperature of sample. Secondly, the  transmittance amplitude of speckles should  be modeled as a stationary random process. Since the corresponding surface area elements are small enough compared to the illumination area on the diffusion screen, consequently, this condition is almost assured for nanoparticles. It turns out that any deviation from above conditions leads to get inaccurate value for speed of nanoparticles. Therefor, we look for  a robust algorithm without changing experimental setup considerably for precise measurement. Suppose that recorded data is represented by $x(t)\equiv I(t)-\langle I(t)\rangle$. The correlation function of the speckle intensity fluctuations produced at the plane of the Fresnel diffraction field under the illumination of the Gaussian beam for arbitrary shape of detector aperture is written by \cite{Asakura,Takai,Asakura81,Takai2,Iwai}:
\begin{eqnarray}
\label{eq29}
&&C_{xx} (\tau)=K \ {\bf e}^{-{v^{2}\tau^{2}}/{\omega^{2}}}\\
&&\times\int d{\bf R} {\mathcal D}({\bf R})\ {\bf exp}{\left(-\left(\frac{\pi \omega}{\lambda \ell}\right)^{2}\left[ {\bf R}-\left( \frac{\ell}{\rho}+1 \right)v\tau\right]^{2}\right)}\nonumber
\end{eqnarray}
where $K$ is a constant, $\omega$ and $\rho$ indicate the width and wavefront curve of the Gaussian beam at the sample position and ${\mathcal D}( {\bf R})$ is the spatial correlation function of the detecting aperture \cite{Iwai}. 
Finally the number of {\it zero-crossing} becomes:
\begin{eqnarray}
\label{eq35}
N_{0}=v\frac{\sqrt{2}}{\pi}\sqrt{\frac{\gamma^2}{\Delta ^2+d^2}+\frac{1}{\omega^{2}}}
\end{eqnarray}
here $\gamma\equiv\frac{\ell}{\rho} +1$ and $\Delta \equiv\frac{\lambda \ell}{\pi \omega}$ representing the average grain size of speckles at the detecting plane. $\ell$ is distance between sample and detector while $d$ is radius of aperture (Fig. \ref{fig:3}). 
Above equation indicates that the number of {\it zero-crossings}
is directly proportional to the average in-plane speed, $v$, of diffuse nanoparticle \cite{Takai}.

As mentioned before, when the detecting aperture is sufficiently large compared to the average grain size of detected speckles, consequently, the probability density function of the spatially integrated speckle-intensity variation has Gaussian form.  In such case, above expression for $N_0$ is a good estimation. In real experimental setup, non-Gaussianity is expected to occur. Therefore, to increase the estimation accuracy of nanoparticles speed, we extend our analysis based on crossing statistics instead of using only {\it zero-crossing} method.

In our experimental setup, we measure the fluctuation of spatially integrated intensity for three different temperatures, namely $50$, $70$ and $90$ $^{\circ}$C. In Tab.  \ref{tab:1}, we report the computed average speed of nanoparticle for various temperature. These values are compatible with those results presented in ref. \cite{Karimzadeh13}. The statistical error is reduced due to more available measurements.  There is a trade off between desired accuracy and computational time consuming. In our analysis, we only take 50 level to compute mean value of particle's speed and finally the statistical errors for relevant quantities reduced by one-order of magnitude. Referring to Eq. (\ref{eq37}), we also determine $N_{total}(0)$. As represented in this table, by increasing temperature of solution, we expect that fluctuations of recorded signal in this experiment to be increased corresponding to higher roughness of series. These fluctuations are directly related to the high diffusive behavior of nanoparticles. It is wroth noting that, without any net flow for suspended particles, we have fluctuations in recorded spatially integrated intensity causing to obtain background level for computed mean-speed value.   In principle, to evaluate the influence of speckle boiling occurring with zero drift velocity at a given temperature, one should put the sample in a heat bath. When sample becomes thermalized, the corresponding values for mean-speed computed by crossing statistics for all available thresholds reveal the background contribution. The mean-speed associated with background is temperature dependent leading to a bias effect. Subsequently, mentioned values should be subtracted from that of computed for the configuration with net heat flow. 
We repeated our measurements for different temperatures without any net heat flow on sample and our results demonstrated that  about 15\% portion of speed value is devoted to background level. To clarify this aspect, we add a column in Tab.  \ref{tab:1} including the value of background level when we consider {\it zero-crossing}. The contribution for other crossing statistics is almost similar. Subsequently, to obtain correct  mean value of particle's speed, we subtract the background level in each threshold and the modified values have been reported in Tab.  \ref{tab:1}.  Hereafter the contribution of background has been subtracted from all given results and plots throughout this paper.

\begin {table*}[ht]
\centering
\caption {The mean value of particle's speed computed in experiments for different temperatures at $1\sigma$ confidence interval. Column representing by "Background level" corresponds to the background measurement when we consider only {\it zero-cross}. In last column the total crossing for $q=0$ has been reported.} \label{tab:1}
\begin{center}
  \begin{tabular}{|c|c|c|c|c|c|  }
    \hline
   Temperature  & $v(0)\pm\sigma_v$&$\langle v(\alpha)\rangle_{\alpha}^{\rm Gaussian}\pm\sigma_v$&$\langle v(\alpha)\rangle_{\alpha}^{\rm non-Gaussian}\pm\sigma_v$   &Background level& $N_{total}(0)$ \\ \hline
      50 $^{\circ}$C &  $2.49 \pm0.10$ $(\frac{mm}{s})$&$2.44\pm0.02$ $(\frac{mm}{s})$ &$2.52\pm0.02 $ $(\frac{mm}{s})$ &$0.44\pm0.10$ $(\frac{mm}{s})$&8238  \\ \hline
     70 $^{\circ}$C &$2.85\pm0.10$ $(\frac{mm}{s})$&$2.76\pm0.02$ $(\frac{mm}{s})$&$2.86\pm0.02$ $(\frac{mm}{s})$&$0.50\pm0.10$ $(\frac{mm}{s})$&10108   \\ \hline
    90 $^{\circ}$C &$3.03\pm0.10$ $(\frac{mm}{s})$&$2.99\pm0.02$ $(\frac{mm}{s})$ &$3.01\pm0.02$ $(\frac{mm}{s})$&$0.54\pm0.10$ $(\frac{mm}{s})$&10822  \\
    \hline
  \end{tabular}
\end{center}
\end{table*}
\begin{figure}[h]
\centering
\includegraphics[width=8cm, height=4cm]{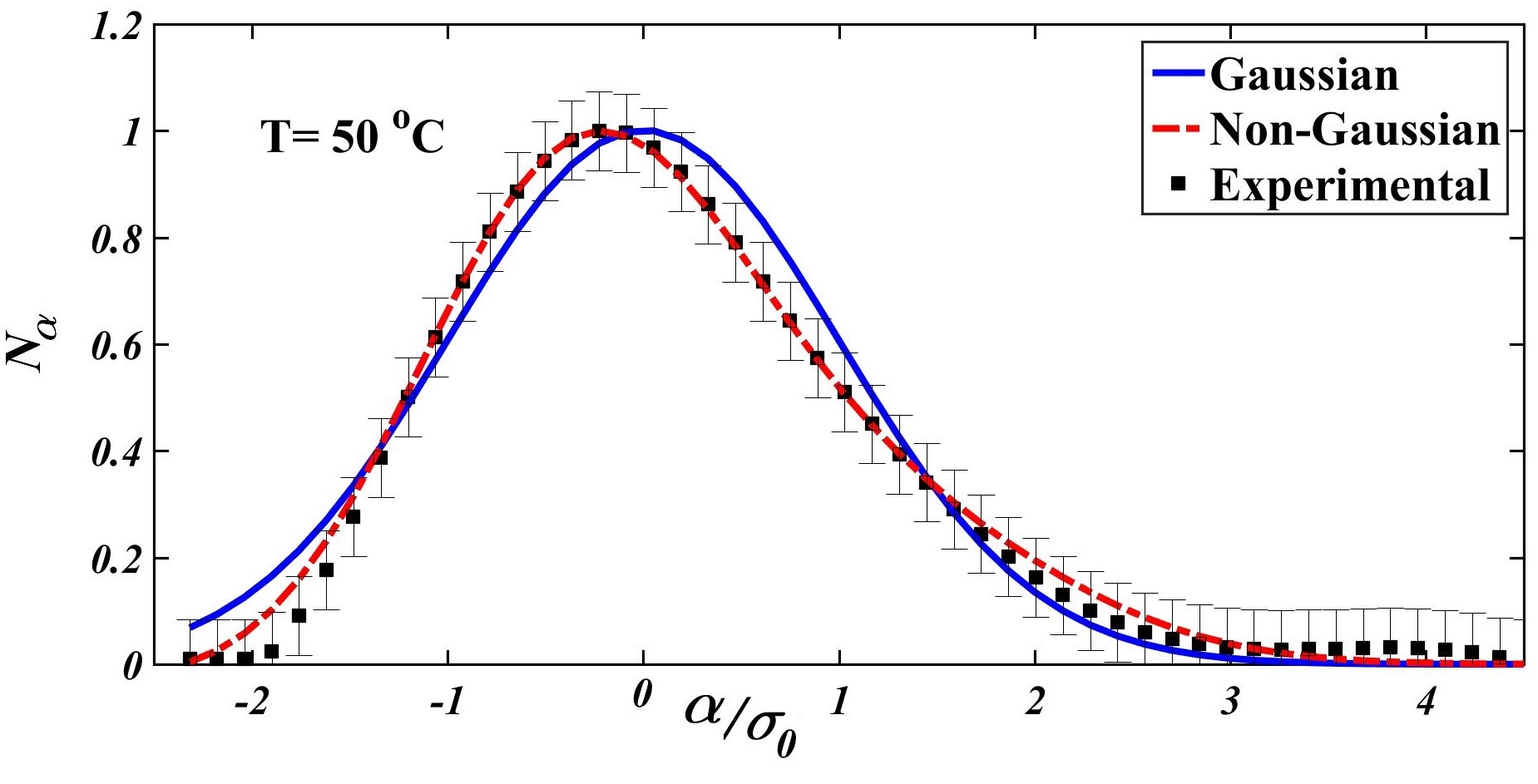}
\includegraphics[width=8cm, height=4cm]{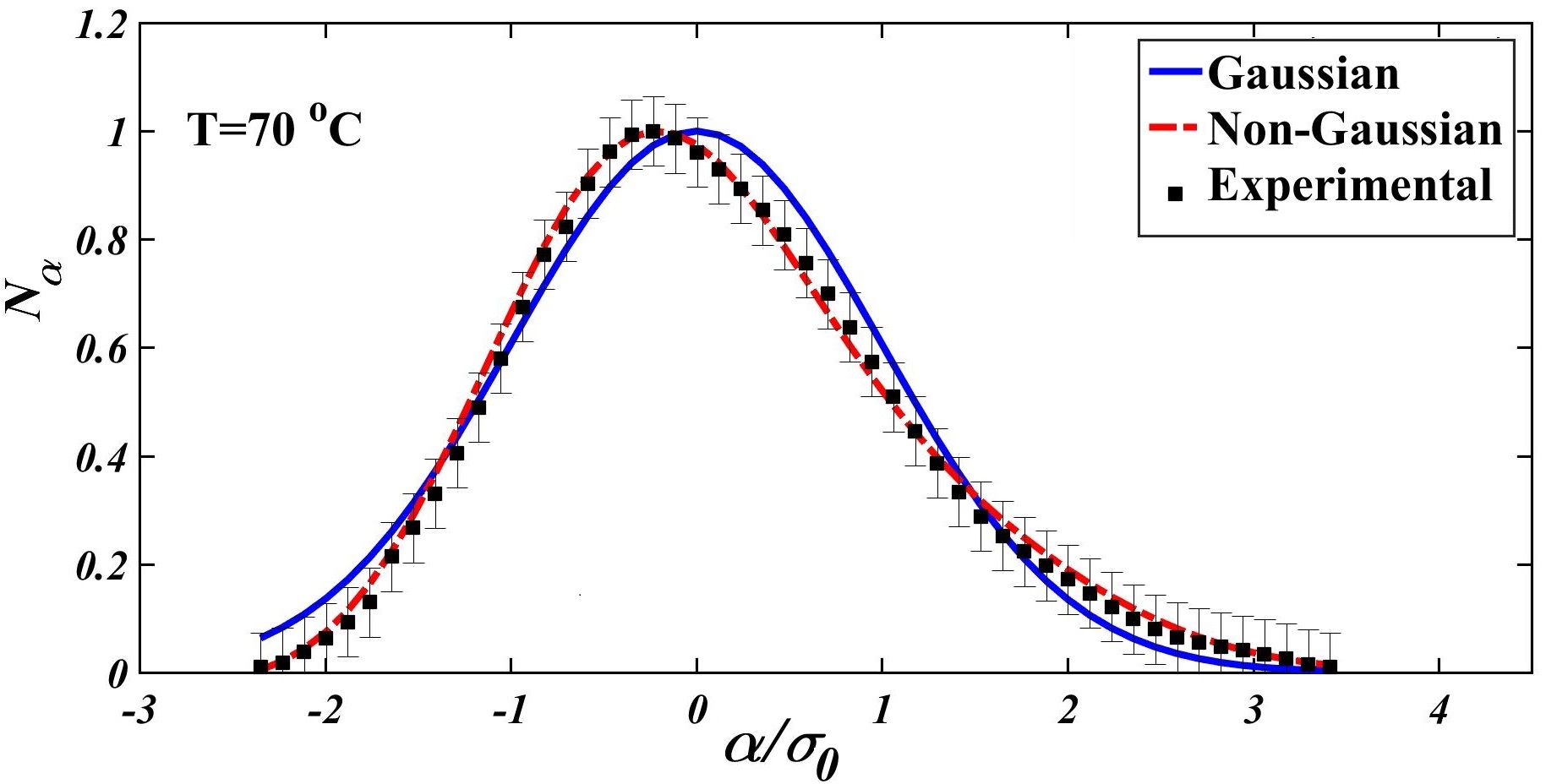}
\includegraphics[width=8.1cm, height=4cm]{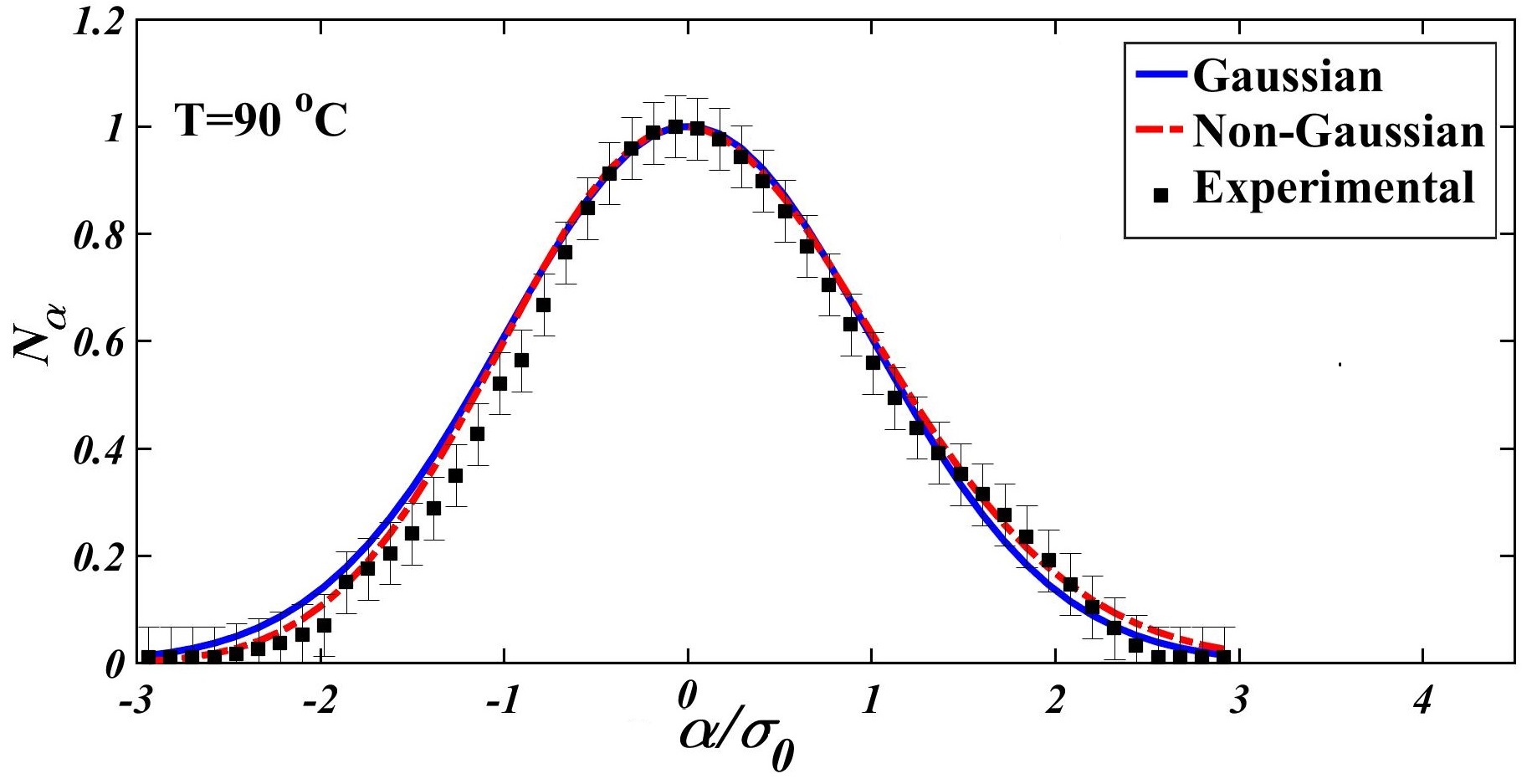}
\caption{Crossing statistics as a function of threshold for our experimental setups for three different temperatures. Upper panel corresponds to $T=50$$^{\circ}$C, middle panel is for $T=70$$^{\circ}$C and lower panel represents crossing statistics for $T=90$$^{\circ}$C. Symbols in these plots indicate $N_{\alpha}$ computed numerically, solid  line shows theoretical prediction of crossing statistics for Gaussian fluctuations. Dashed line represents $N_{\alpha}$ in the context of perturbation theory up to $\mathcal{O}(\sigma_0^2)$ (Eq. (\ref{pertur1})).
}
\label{fig:6}
\end{figure}

So far, we could compute the value of mean-speed of nano-particles suspended in fluid, but the crossing statics prepares more complete approach to reduce statistical errors arising in experiment. In addition to {\it zero-crossing} method, we compute crossing statistics for all available values of threshold and by stacking all results with own weighted errors, we are able to estimate more precise values for relevant physical quantities such as mean value of nanoparticles speed. 

The crossing statistics in terms of threshold for three experiments is represented in Fig. \ref{fig:6}. Symbols indicates  $N_{\alpha}$ versus $\alpha$ computed numerically, while solid line is devoted to theoretical prediction for Gaussian fluctuations. Dashed line is associated with non-Gaussian theory.  These plots demonstrate that our experimental setup produces non-Gaussian fluctuations converging to Gaussian when sample temperature to be increased. This result is compatible with our expectation, because, more randomness leads to have more Gaussian fluctuations. To clarify the source of non-Gaussian behavior of spatially integrated fluctuations which can be thought as combination of different reasons during the light scattering of laser light through sample, we should utilize different layouts for experimental setup. Light scattering includes an average over all fluctuations in both space and time. Therefore, statistical mechanics provides relation between the intensity and frequency distribution of the scattered light to the thermodynamic and transport properties of the fluid \cite{moun66,gold99}. In principle, following reasonable effects can be considered to describe the non-Gaussian statistics of  probability density function of light scattering fluctuating intensity:  non-homogeneity of particle distribution in the solution and in scattering volume, non-uniformly temperature distribution, temperature dependency of cluster size and mass in solution, wavelength of incident light, polarizability, interaction between particles at different temperatures, absorption factor, viscosity, pair correlation function and so on.  Generally, scattered intensity arriving at the detector has two parts. First part is devoted to slow fluctuations due to variations in the total number of scatterers in the scattering volume. While,  the second part corresponds to interference fluctuations due to superposition of randomly fluctuating phases \cite{scha72,bern74}. The characteristic time scale of first part corresponds to average time that it takes the scatterers to cross scattering volume. Whereas, the time scale for second part is given by diffusion contribution. The number fluctuations provide slowly varying behavior in fluctuating intensity and it belongs to the deterministic part of particles movement, consequently, sharp peaks in intensity is mostly affected by interference fluctuations which is represented by stochastic portion of particle's motion. Indeed, the higher value of temperature, the higher rate of multiple scattering leading to more Gaussian fluctuations due to central limit theorem \cite{scha72,bern74}. The fluctuation intensity of light scattering depends on structure factor including the pair correlation function. Lower temperature makes more long-range ordered behavior, while higher temperature corresponds to more randomized particles due to increasing diffusion coefficient of the particles. Consequently, at lower temperature, pair correlation function has more features which deviates from unity leading to have more sharp values in light intensity. In other word, higher temperature washes out any excess probability in finding particles with respect to random case \cite{pathria11}.  The deterministic part of particle's movement due to net flow mainly affects on the overall value of light scattering intensity and higher value of net movement decreases the overall value of detected intensity.

\begin{figure}[h]
\centering
\includegraphics[width=8.1cm, height=3.5cm]{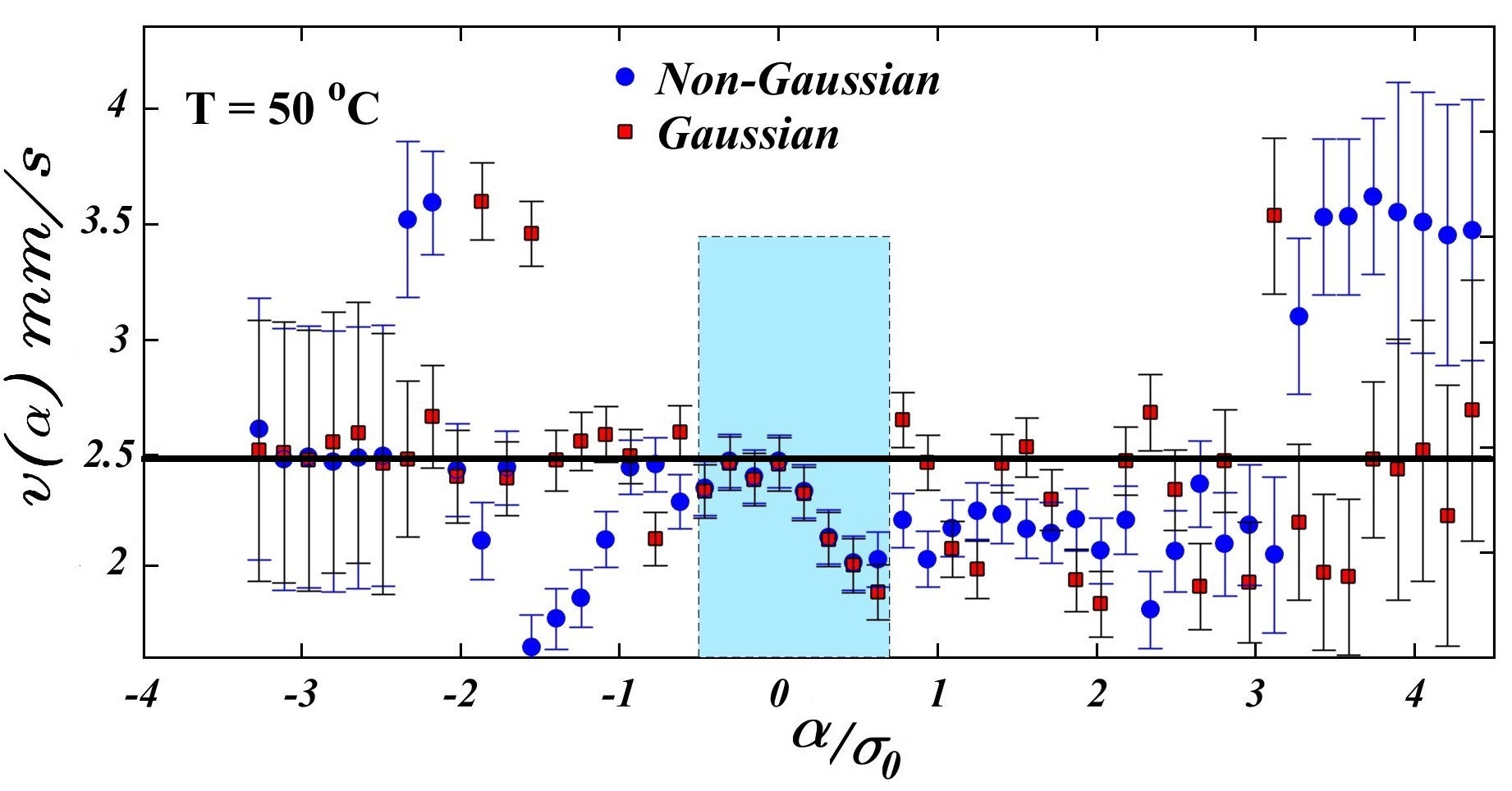}
\includegraphics[width=8.1cm, height=3.5cm]{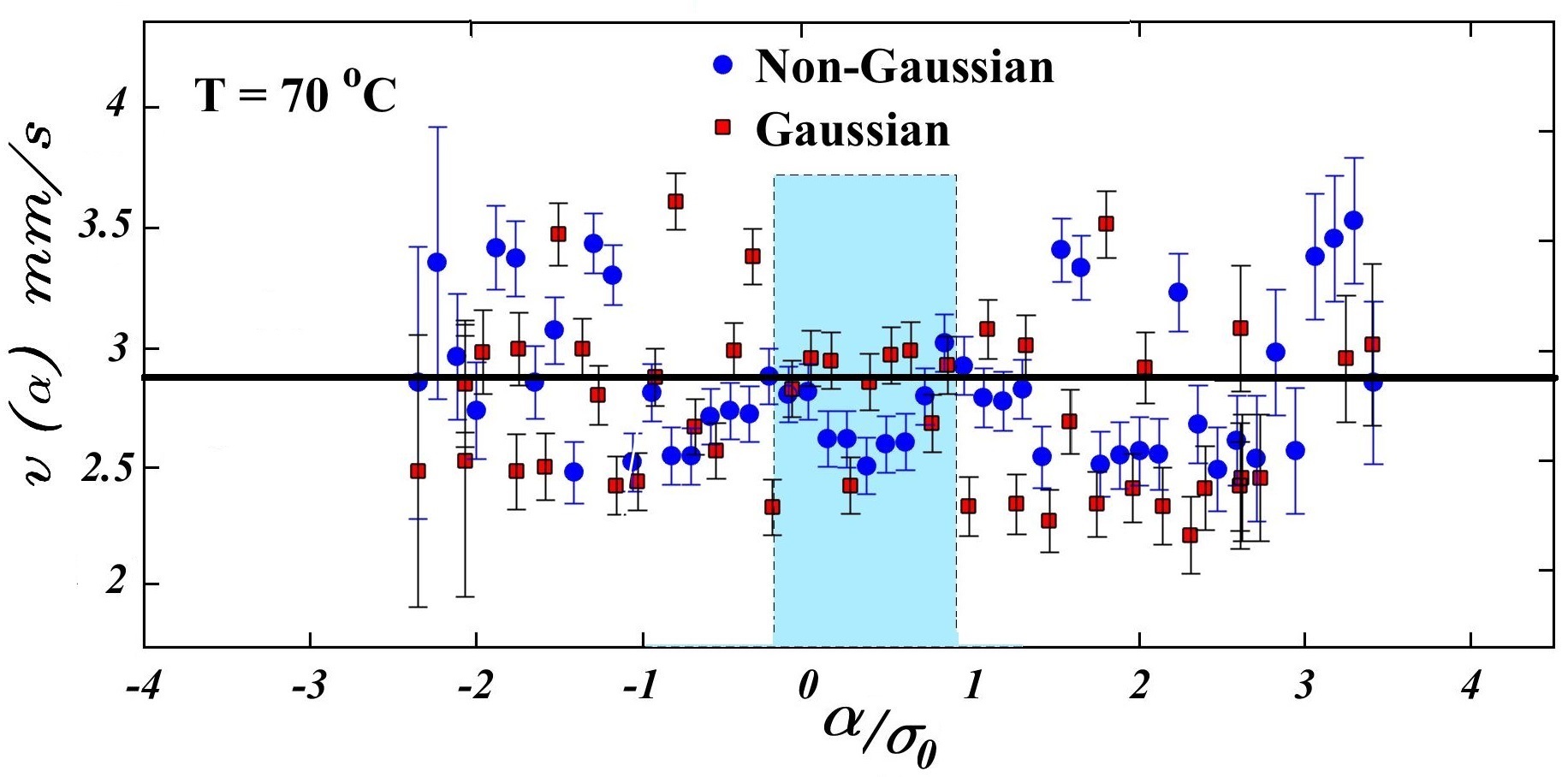}
\includegraphics[width=8.25cm, height=3.5cm]{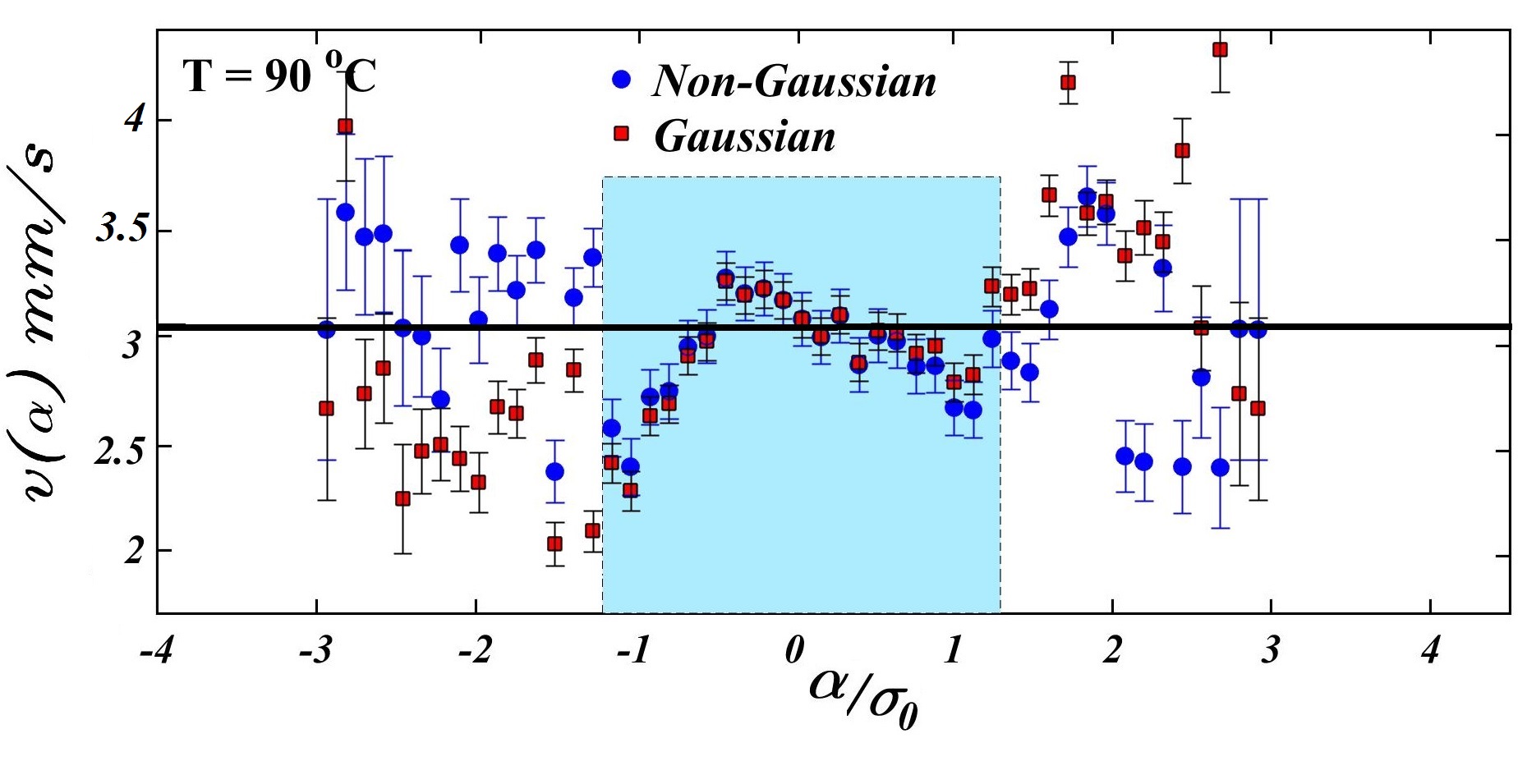}
\caption{Measured mean value of particle's  speed for three temperatures based on various levels. In each panel, filled square symbols are devoted to Gaussian theory while circle symbols represent results based on non-Gaussian perturbative theory up to $\mathcal{O}(\sigma_0^2)$. Horizontal solid line in each plot represents the particle's speed computed by $zero$ level. All values in this plots have been modified by subtracting background level. }
\label{fig:9}
\end{figure}

According to theoretical prediction for crossing statistics  at a given threshold using Gaussian theory, Eq. (\ref{eq27}), we compute mean value of nanoparticle speed.  The quantity $\langle v(\alpha)\rangle_{\alpha}$ corresponds to the computed mean-speed averaged on all available levels. These values are in agreement with that of determined by only $zero-crossing$ method. Statistical dispersion of mean-speed, $\sigma_v^2=\langle (v(\alpha)-\langle v(\alpha)\rangle)^2\rangle_{\alpha}$, is reported in Tab. \ref{tab:1}. 
Taking into account the non-Gaussianity of fluctuations leads to more precise value for $v$ (Eq. (\ref{pertur1})) as reported in Tab. \ref{tab:1}. In Fig. \ref{fig:9} we indicate the computed mean value of particle's speed for each level. Our results demonstrate that   the consistency interval for threshold between Gaussian and non-Gaussian theory for computing relevant quantity is less than $2\sigma$ centered by mean value for low temperature. Mentioned interval increases beyond $2\sigma$ for higher temperature. Subsequently for computing mean value of speed, the weak non-Gaussian model should be considered in order to achieve precise values.

\begin{figure}[h]
\centering
\includegraphics[width=8.1cm, height=4cm]{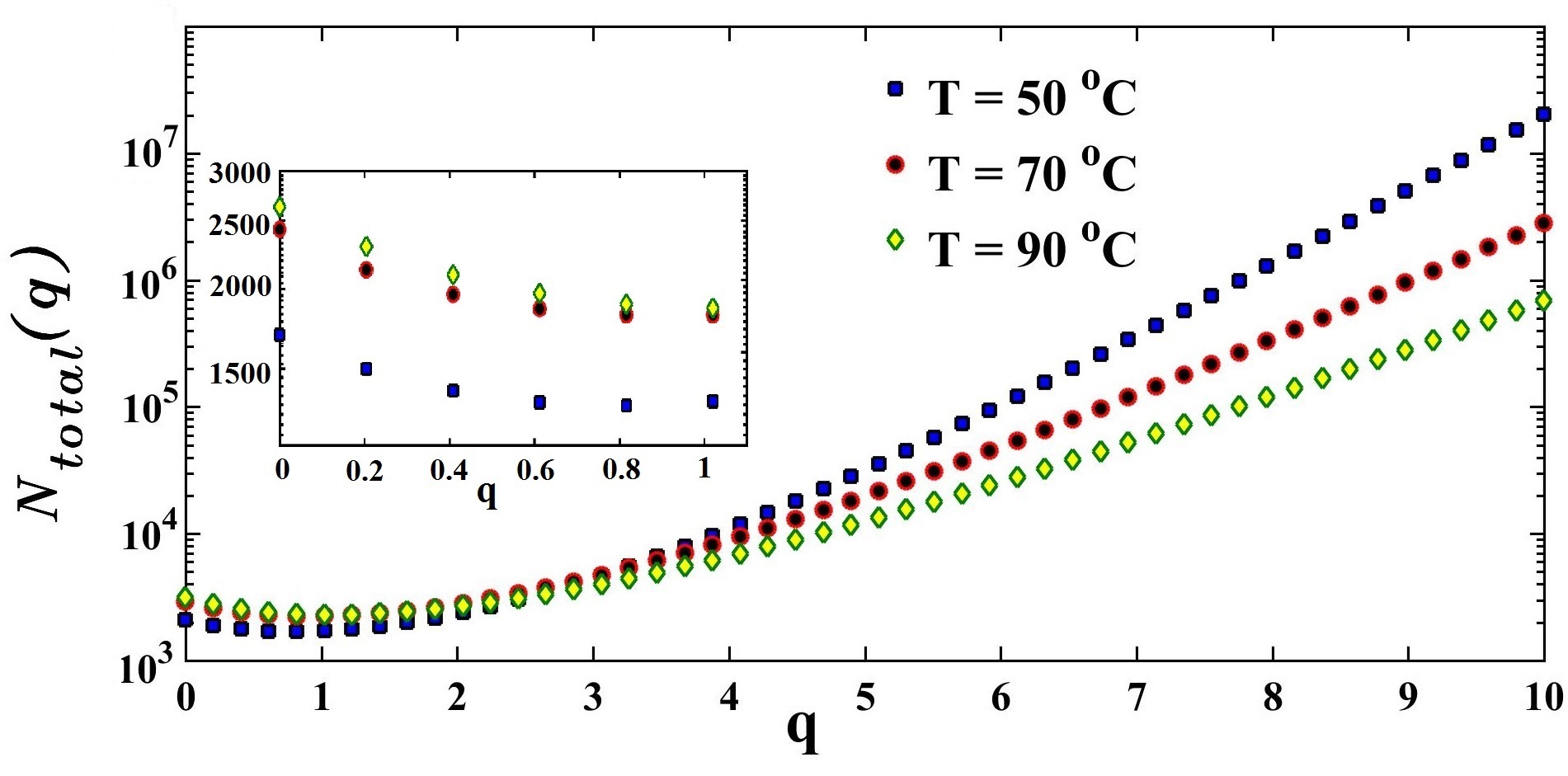}
\caption{Numerical value of $N_{total}$ in the semi-log curve versus $q$ for experimental data. Triangles, squares and circles are related to results for $T=50^{\circ}$C, $T=70^{\circ}$C and $T=90^{\circ}$C, respectively. The inset plot is a magnification for small value of $q$ to make more sense. All values in this plots have been modified by subtracting background level. }
\label{fig:11}
\end{figure}

\begin{figure}[h]
\centering
\includegraphics[width=8.1cm, height=3.8cm]{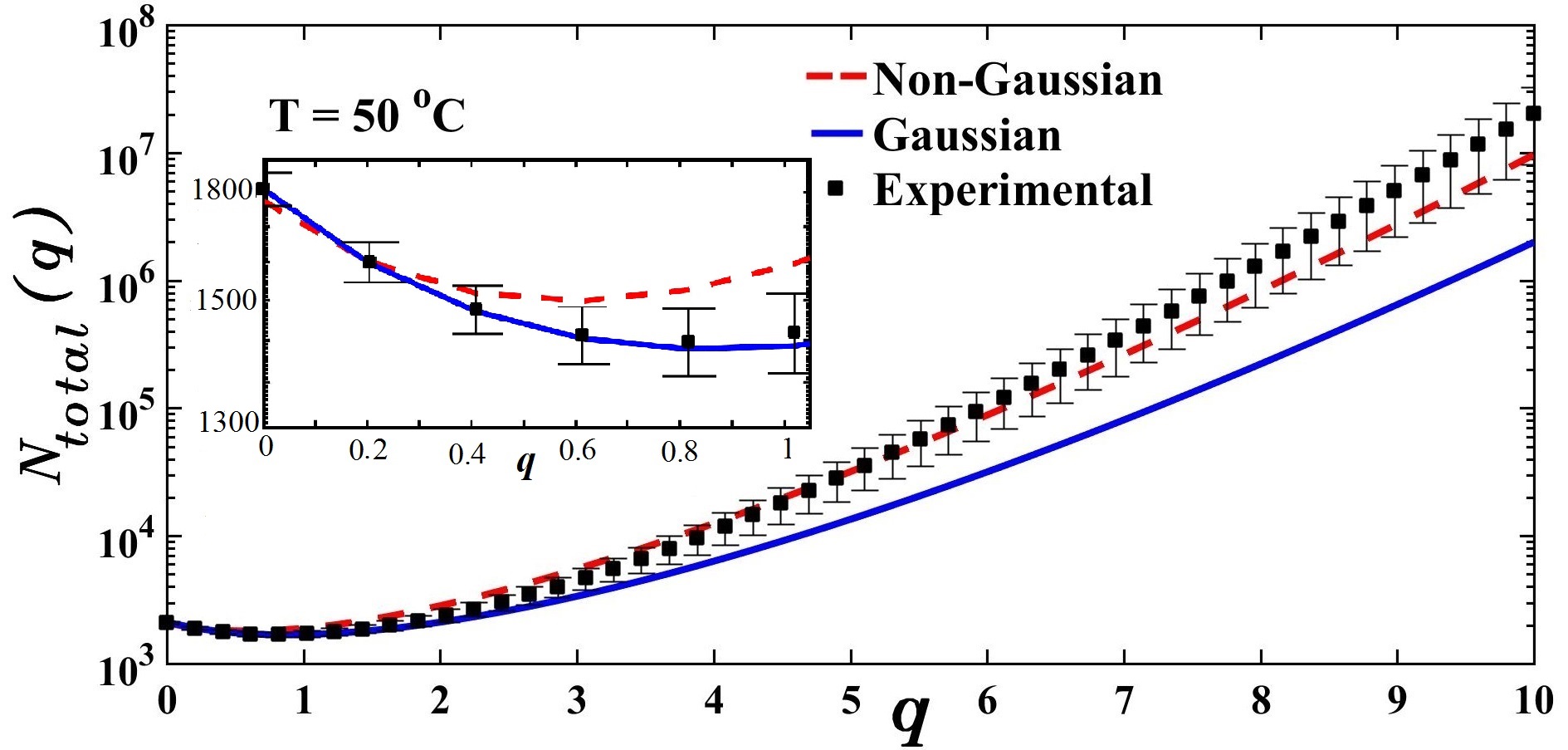}
\includegraphics[width=8.1cm, height=3.8cm]{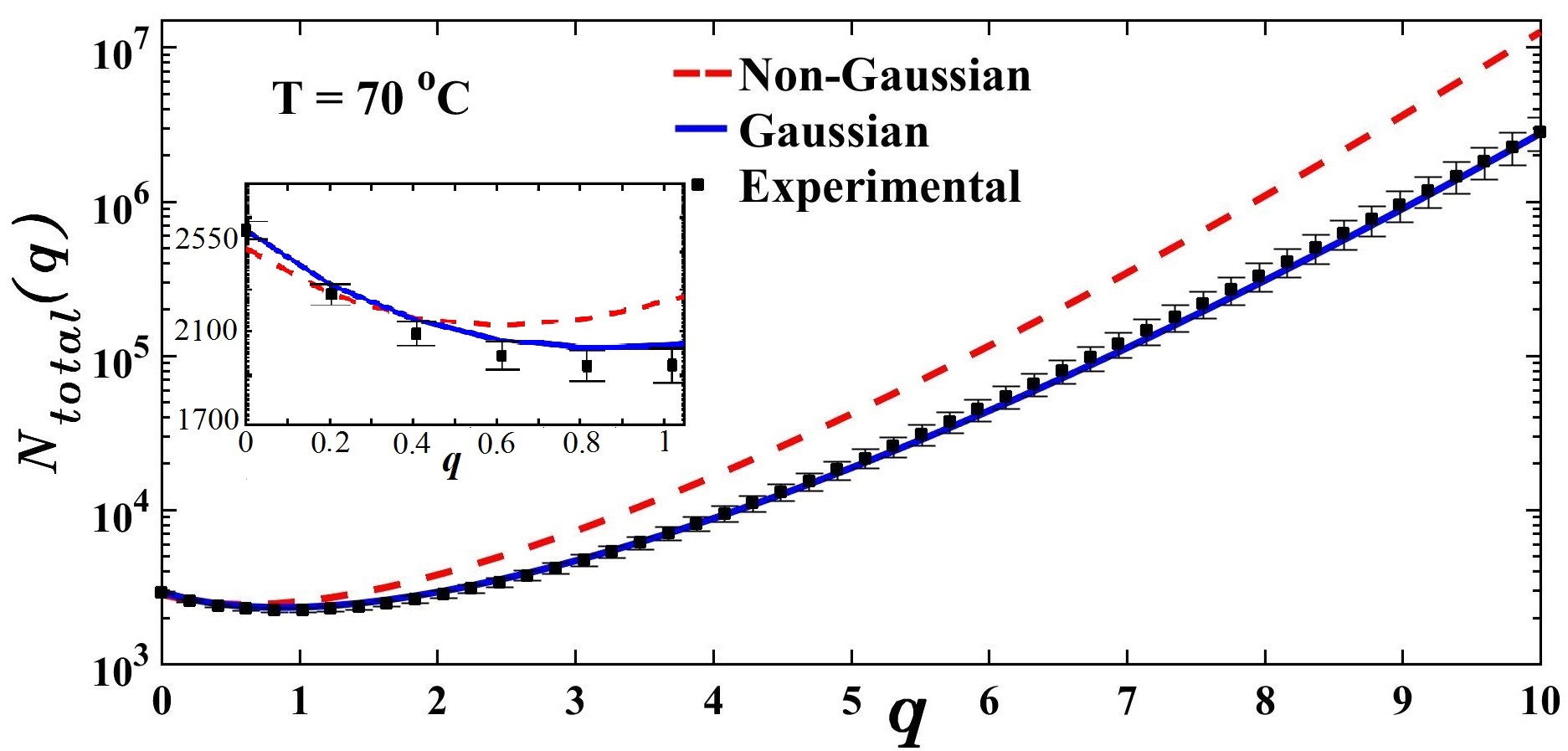}
\includegraphics[width=8.1cm, height=3.8cm]{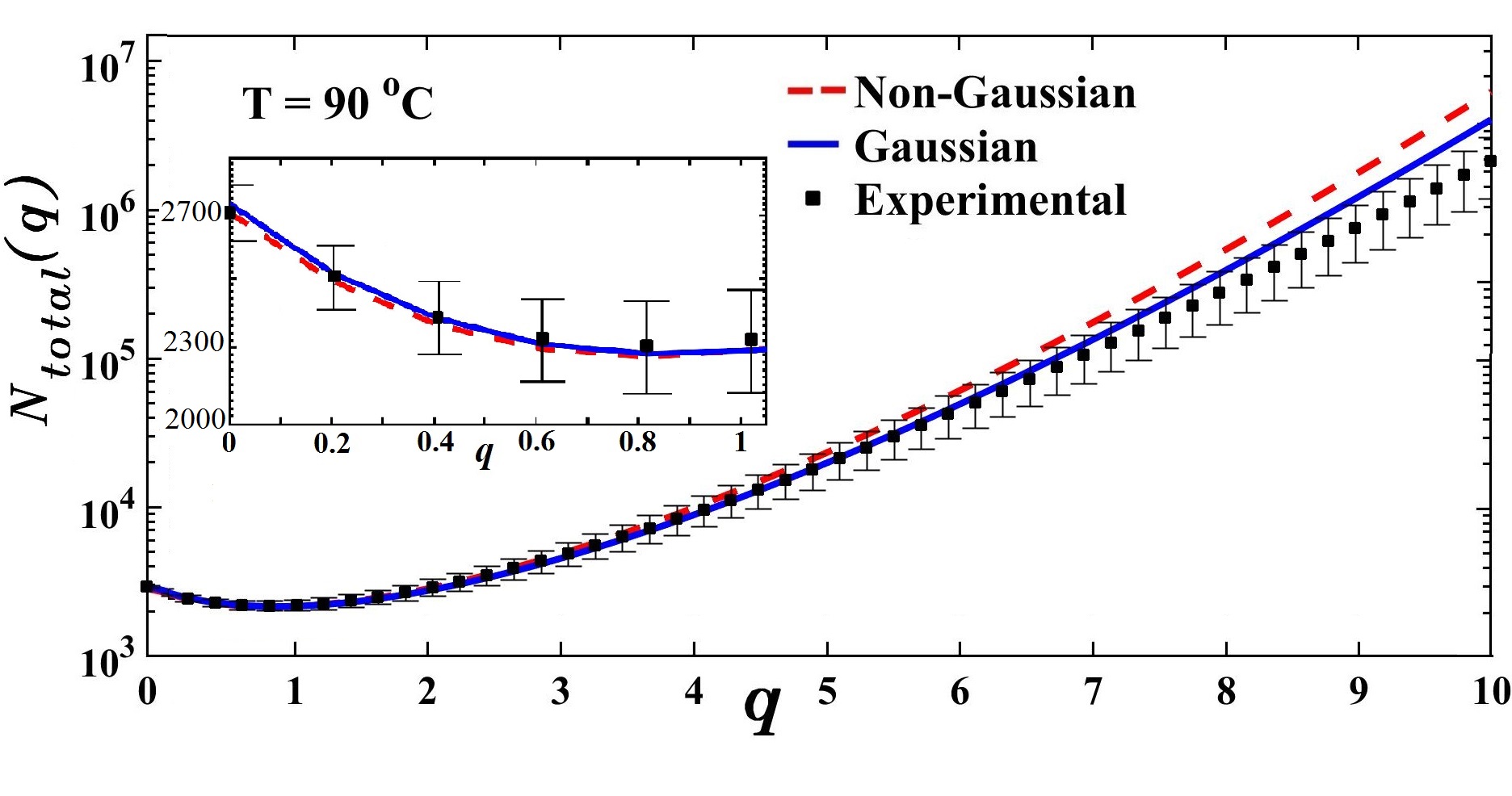}
\caption{Semi-log plot of $N_{total}$ versus $q$ for different sample temperatures. Filled symbols are devoted to numerical computation. Dashed line corresponds to perturbative non-Gaussian theory up to $\mathcal{O}(\sigma_0^2)$ while solid line indicates result for Gaussian theory. The inset plot is a magnification for small value of $q$ to make more sense. All values in this plots have been modified by subtracting background level.}
\label{fig:10}
\end{figure}

\begin {table}
\centering
\caption {The value of threshold for which we get maximum value of crossing ($N_{\alpha}^{max}$) and corresponding characteristic time scale for three experiments.} \label{tab:4}
\begin{center}
  \begin{tabular}{|c|c|c| }
    \hline
   Temperature  & $\alpha/\sigma_0$&$\tau_{\alpha}$ (sec) \\ \hline
      50 $^{\circ}$C &$-0.23$  &$0.15\pm0.01$ \\ \hline
     70 $^{\circ}$C &$-0.23$&$0.13\pm0.01$   \\ \hline
    90 $^{\circ}$C &$-0.07$&$0.12\pm0.01$   \\
    \hline
  \end{tabular}
\end{center}
\end{table}

Now we focus on generalized total number of crossing statistics introduced in Eqs. (\ref{eq277}) and (26). $N_{total}(q)$ for $q=0$ characterizes the total number of fluctuations for all thresholds. This value is directly devoted to roughness of underlying fluctuations. As we expect, higher sample temperature corresponds to higher value of $N_{total}(q=0)$ (see Fig. \ref{fig:11}).  The higher value of slope for large $q$ clarifies the higher probability of having large fluctuations in intensity of light fluctuations detected during the measurement. Fluctuations in light passing through the sample in lower temperature are highly encoded with spike-shape due to deviation from Gaussian assumptions as explained before, 
while by increasing the sample temperature, the  rate of multiple scattering grows up. 
Subsequently,  the probability of finding the large fluctuations in beam intensity becomes small as seen in Figs. \ref{fig:4} (lower panel) and \ref{fig:6}. In Fig. \ref{fig:10},  we plot $N_{total}$ as a function of $q$ for three samples. In this plot symbols correspond to numerical computation of generalized total crossings. Solid line predicted by Eq. (26) while Gaussian model (Eq. (\ref{eq277})) is illustrated  by solid line. This measure is very sensitive to existence of rare events. For higher speed of particle, we expect to have more particle collision causing the intensity of speckles varies rapidly, so the average of speckles intensity fluctuations is almost concentrated around the value of mean fluctuations. Therefore,  the slope of $N_{total}(q)$ for large value of $q$ becomes smaller than that of given for low temperature.
Using the $N_{total}(q)$ quantity, it is possible to examine the contribution of various size of fluctuations. For $q<1$,  small fluctuations have dominant impact, on the contrary, for $q\ge1$ large fluctuations play main role in computing generalized crossing statistics. In our experimental setup, the well-known Gaussian theory can not reveal proper view for mean value of speed of suspended particles in solution which is affected by thermal fluctuations and other relevant phenomena. 
On the contrary,  to achieve more precise characterization of underlying sample,  perturbative crossing statistical approach introduced in context of statistical analysis of random fields \cite{matsubara03} is an alternative approach. According to Eqs. (26) and (\ref{eq35}), the corresponding mean-speed for different $q$ can be written as:
\begin{widetext}
\begin{eqnarray}
\label{pertur2}
v(q)&=& \frac{N^{\rm exp.}_{total}(q)}{\beta(2\sigma_0^2)^{(q+1)/2} \Gamma\left(\frac{q+1}{2}\right)}\left\{ 1-\sigma_0 \left[\frac{(2+q)\sigma_0^2-3}{6}S^{(0)}+\frac{S^{(1)}}{3}\right]\frac{\frac{q}{2}\Gamma\left(\frac{q}{2}\right)}{\Gamma\left(\frac{q+1}{2}\right)}+\mathcal{O}(\sigma_0^2)\right\}
\end{eqnarray}
\end{widetext}
here $\beta=\frac{\sqrt{2}}{\pi}\sqrt{\frac{\gamma^2}{\Delta^2+d^2}+\frac{1}{\omega^2}}$ and $N^{\rm exp.}_{total}(q)$ is determined directly in experiment.  Taking into account smaller value for $q$ leads to reduce contribution of spurious fluctuations with large values in intensity for computing mean-speed, while for artificial effects in range of small fluctuations, larger value of $q$ becomes more reasonable to employ.

The average time scale that the speckle intensity remains above a given threshold on each crossing has been introduced in \cite{Pratt73}. Inspired by mentioned quantity, one can define an interesting measure based on crossing statistics which is characteristic time scale for having statistical successive crossings at a given threshold. To this end, one can define $\tau_{\alpha}\equiv1/N_{\alpha}$. According to mentioned characteristic time scale, our results show that the minimum value of time interval for having crossing corresponds to maximum value of $N_{\alpha}$ achieves for threshold equates to $\alpha/\sigma_0=-0.23$ for $T=50^{\circ}$C, $\alpha/\sigma_0=-0.23$ for $T=70^{\circ}$C and $\alpha/\sigma_0=-0.07$ for $T=90^{\circ}$C. These values have been reported in Tab. \ref{tab:4}. Higher temperature leads to more coincidence between peaks of crossing statistics curve determined by different approaches. This means that $zero$
 threshold for only high enough temperature can support reliable physical properties.

\section{Results and discussion}\label{result}
The stochasticity nature of spatially integrated speckle
intensity fluctuations encoded in scattered laser light leads to look for robust statistical method rather than common approach in analyzing underlying fluctuations. In this paper, we extend  the particular case of crossing statistics, namely {\it zero-crossing} to measure the mean-speed of moving particles. The crossing events not only at {\it zero} level but also for all of available thresholds are considered for further calculations. 

As discussed in more details in section~2, we revisited the up- and down-crossing statistics for both Gaussian and non-Gaussian processes. The generalized form for crossing statistics as a function of threshold, $N_{\alpha}$ (Eq. (\ref{pertur1})),  and total crossing, $N_{total}(q)$ (Eq. (26)), have been demonstrated. Using proper experimental setup, we measure the fluctuations of spatial integrated intensity of scattered laser light at different temperatures, namely 50, 70 and 90 $^{\circ}$C. We represent the recorded fluctuation as a $1+1$-Dimensional stochastic field by $\{x(t)\}\equiv\{I(t)-\langle I(t)\rangle \}$.  For mentioned stochastic field, we compute the number of {\it zero-crossing} statistics and by using Eq. (\ref{eq35}), the mean value of speed of nanoparticles is determined. In the left second column of Tab.~\ref{tab:1}, we report the mean value of particles's speed with corresponding errors at $1\sigma$ confidence interval. 

Fig. \ref{fig:6} indicates  $N_{\alpha}$ versus $\alpha$ computed numerically (symbols). In this plot solid line is devoted to Gaussian theoretical prediction. Dashed line is associated with non-Gaussian theory. These plots depict our experimental setup produces non-Gaussian fluctuations converging to Gaussian when sample temperature to be increased. To get higher degree of accuracy for mean-speed of particles, we take into account all of available thresholds based on Gaussian and non-Gaussian theories. The value of $\langle v(\alpha)\rangle_{\alpha}$ and corresponding value of statistical dispersion, $\sigma_v^2$, are reported in Tab.~\ref{tab:1}.  Referring to Fig. \ref{fig:6}, it is elucidated that for lower temperature, the probability to achieve crossing at higher threshold is considerable compared to higher temperature. Scattered intensity arriving at the detector has two parts: first part is devoted to slow fluctuations due to variations in the total number of scatterers in the scattering volume, while,  second part corresponds to interference fluctuations due to superposition of randomly fluctuating phases \cite{scha72,bern74}. The number fluctuations provide slowly varying behavior in fluctuating intensity and it belongs to deterministic part of particles movement, consequently, sharp peaks in intensity is mostly affected by interference fluctuations which is represented by stochastic portion of particle's motion. Namely, the higher value of temperature the higher rate of multiple scattering leading to more Gaussian fluctuations due to central limit theorem \cite{scha72,bern74}. From structure factor point of view, we can express that, for lower temperature, pair correlation function has more features which deviates from unity leading to have more sharp values in light intensity. The deterministic portion of particle's movement due to net flow mainly affects on the overall value of light scattering intensity and higher value of net movement decreases the overall value of detected intensity (see Fig. \ref{fig:4}). 

The computed mean value of particle's speed for each level is indicated in Fig. \ref{fig:9}. The consistency interval for threshold between Gaussian and non-Gaussian theory for mean value of particle's speed is less than $2\sigma$ centered by mean value for low temperature. Mentioned interval increases beyond $2\sigma$ for higher temperature. As we expect, for computing the speed mean value, we should utilize the weak non-Gaussian model.

Generalized total number of crossing statistics introduced in Eqs. (\ref{eq277}) and (26) specify the weighted total number of fluctuations for all thresholds. As an illustration, for $q=0$, this value is directly devoted to roughness of underlying fluctuations. The higher sample temperature corresponds to higher value of $N_{total}(q=0)$ as indicated in  Fig. \ref{fig:11}.  The $N_{total}$ as a function of $q$ for three samples is plotted in Fig. \ref{fig:10}.  In this plot symbols correspond to numerical computation of generalized total crossings. Solid line predicted by Eq. (26) while Gaussian model (Eq. (\ref{eq277})) is illustrated  by solid line. The higher value of slope for large $q$ clarifies the higher probability of having large fluctuations in underlying process. By increasing the sample temperature, the convection speed becomes dominant in comparison with other effects. Subsequently,  the probability of finding the large fluctuations in beam intensity becomes small as seen in Figs. \ref{fig:4} (lower panel) and \ref{fig:6}. This measure is very sensitive to existence of rare events. At high temperature the slope of $N_{total}(q)$ for large value of $q$ becomes smaller than that of given for lower temperature. In our experimental setup, the well-known Gaussian theory can not reveal proper view for relevant physical quantities.  To quantify the value of deviation from Gaussian theory, we computed $N_{total}$ as a function of $q$ indicated in  Fig. \ref{fig:10}. For larger value of $q$, higher fluctuations have dominant contribution, on the contrary for $q<1$, small fluctuations play predominant role. We found that by increasing sample temperature, higher value of fluctuations are almost suppressed. Also, the non-Gaussian terms vanish and non-Gaussian model as well as experimental data  converge to Gaussian situation.  

Interestingly, according to Eqs. (\ref{pertur1}) and (\ref{eq35}) it is possible to compute mean-speed for different $q$ which is represented by Eq. ~(\ref{pertur2}). Smaller value of $q$ reduces contribution of spurious large fluctuations, while for artificial effects in range of smaller value of fluctuations, larger value of $q$ becomes more reasonable to employ.     

All quantities determined in our experimental setup are induced by background effect and in order to find correct values for mean-speed and other values,  we should subtract this effect in all measurements.

Characteristic time scale for having statistical successive crossings at a given threshold defined by $\tau_{\alpha}\equiv1/N_{\alpha}$ has been reported in Tab. \ref{tab:4}. Our results demonstrate that lower temperature leads to higher deviation between the level with maximum value of crossing and {\it zero} level. This finding confirms that using only $zero$ threshold for high enough temperature is able to support precise physical properties

\section{Conclusion}\label{summary}

We can extract viable physical quantities concerning dynamical properties of suspended particles by utilizing various statistical analysis of recorded fluctuations of scattered laser beam through a nanofluid. 
 Computing PDF and correlation function of  scattered laser beam fluctuations through a nanofluid can reveal the size, shape, clustering of suspended particles and type of interaction \cite{Asakura81,gold99,Pratt73,brar11,braun11,chic08,boyd11}. Meanwhile, a proper method to determine mean value of particle's speed is crossing statistics. A particular case of crossing statistics, namely {\it zero-crossing} was carried out to measure the mean-speed of moving particles \cite{Asakura,Takai, Iwai,Asakura81, Takai2,Barakat}. This model contains some priors and assumptions. In principle, there is no grantee to satisfy mentioned presumptions in a real experimental setup leading to inaccurate measurements.

In this paper by measuring the spatially integrated speckle
intensity fluctuations encoded in scattered laser light, we extend
{\it zero-crossing} analysis and applied crossing statistics
incorporating crossing for all available thresholds in Gaussian and
weakly non-Gaussian regimes. This modification prepares following
advantages in this regards: {\bf Firstly}, the crossing statistics not
only at mean level but also for any arbitrary thresholds is used for
computing more precise dynamical quantities without changing the
experimental setup, considerably. {\bf Secondly}, Gaussianity and
non-Gaussianity of underlying process can be examined systematically.
{\bf In addition}, according to perturbation approach,  challenges for determining JPDF is replaced by deriving an expansion formula for arbitrary statistical measure if the second cumulant to be finite.  

We also introduced generalized total crossing statistics. If fluctuations with larger values are affected by artifacts due to experimental setup and etc. more than smaller fluctuations, therefore considering smaller values of $q$ can reduce spurious results in determining mean-speed. On the contrary, larger values of $q$ is suitable to magnify the contribution of fluctuations far from the mean value, statistically (Eq. (\ref{pertur2})).     
Finally according to crossing statistics, some characteristic time
scales which probably contain physical properties enabling us to
achieve deep insight associated with underlying sample, are
introduced.

We determined crossing statistics of laser light fluctuations for three different sample temperatures. 
  Our results demonstrated that by increasing sample temperature, experimental results, Gaussian and non-Gaussian are consistent together around mean level. 
We found that for low temperature the consistency interval for threshold between Gaussian and non-Gaussian theory for computing relevant quantity is less than $2\sigma$ centered by mean value. Mentioned interval increases beyond $2\sigma$ for higher temperature. Subsequently for computing mean value of speed, the weak non-Gaussian model should be considered in order to achieve robust results. 
Obviously when we consider all available thresholds, the statistical uncertainty for mean value of speed decreases. For all cases, the mean value of speed determined by weakly non-Gaussian theory is higher than that of computed by Gaussian framework.
In our experimental setup, background effect modified the crossing statistics of laser light fluctuations. To achieve reliable results this contribution should be subtracted. 

Characteristic time scale using crossing statistics was defined by inverse value of $N_{\alpha}$ for a given threshold. If the maximum value of crossing happens at $zero$ level, therefore, the minimum value of time interval for having crossing ($\tau_0=1/N_0$) corresponds to $N_{0}^{-1}$. While for our results, the associated threshold for minimum time scale was deviated from $zero$ level. 
 
However, in our experimental setup, the computed mean-speed of diffuse objects according to Gaussian and non-Gaussian model are statistically consistent, but it turns out that  this consistency is not essentially satisfied for other range of temperatures and different concentrations.

It could be interesting to apply this method for different experimental configurations to asses the capability of this modification in crossing statistics for determining mean-speed and its dependency to particle size and viscosities of solution.  Also based on directionality nature of crossing statistics,  this method enables us to examine statistical isotropy of intensity fluctuations which is of interest \cite{Ghasemi15} and it is outside the scope of present study.\\

\section{acknowledgments}
SMSM  was partially supported by a grant of deputy for researches and technology of Shahid Beheshti University. SMSM  also appreciates the school of Physics, Institute for Researches in Fundamental Sciences (IPM), where some parts of this work have been done.


\begin{thebibliography}{1}

\bibitem{adler81} R. J. Adler, {\em The Geometry of Random fields}, (John Wiley and Sons, 1981).
\bibitem{adler07} R. J. Adler and J. Taylor, {\em Random Fields and Geometry}, (Springer, 2007).

\bibitem{rice44}
 S. O. Rice, " Mathematical Analysis of Random Noise," Bell System Tech. J.  {\bf 23}, 282-332  (1944); Bell System Tech. J. {\bf24}, 46 (1945).
\bibitem{percy00}
P. H. Brill, "A Brief Outline of the Level Crossing Method in Stochastic Models" CORS Bulletin {\bf34}, 4 (2000).

\bibitem{jafari06}
 G. R. Jafari, M. S. Movahed, S. M. Fazeli, M. R. Rahimi Tabar, S. F. Masoudi, "Level crossing analysis of the stock markets," J. Stat. Mech. {\bf 06}, P06008  (2006).
 
\bibitem{Vahabi}
M. Vahabi, G. R. Jafari and S. M. S. Movahed, "Analysis of fractional Gaussian noises using level crossing method"  j. stat. mech. {\bf 11}   p11021 (2011).

\bibitem{newland}
D. E. Newland, {\em An introduction to Random vibrations, spectral and wavelet analysis}, Third Edition, (Longman Scientific Technical, 1993).
\bibitem{Khimenko77}
V. I. Khimenko, "The average number of trajectory overshoots of a non-Gaussian random process above a given level," Izv. Vyssh. Uchebn. Zaved., Radiofiz. {\bf 21}, 1170-1176 (1977).
 

\bibitem{shahbazi03}
F. Shahbazi, S. Sobhanian, M. Reza Rahimi Tabar, S. Khorram, G. R. Frootan and H. Zahed, "Level crossing analysis of growing surfaces," J. Phys. A: Math. Gen. 36, 2517-2524 (2003).


\bibitem{burger06}
 M. Sadegh Movahed, A. Bahraminasab, H. Rezazadeh and A. A. Masoudi, "Level Crossing Analysis of Burgers Equation in 1+1 Dimensions," J. Phys. A: Math. Gen. {\bf39}, 3903-3909 (2006).

\bibitem{Ghasemi15}
 M. Ghasemi Nezhadhaghighi, S. M. S. Movahed, T. Yasseri, S. M. Vaez Allaei, "Crossing Statistics of Anisotropic Stochastic Surface,"  {\bf arXiv} 1508.01409 (2016).

\bibitem{ebel79}
 K. Ebeling, "Experimental investigation of some statistical properties of monochromatic speckle patterns," Opt. Acta {\bf 26}, 1505-1521 (1979).

\bibitem{barakat80}
 R. Barakat, "The level-crossing rate and above-level duration time of the intensity of a Gaussian random process," Inf. Sci. (N.Y.) {\bf 20}, 83-87 (1980).

\bibitem{bahu80}
 R. D. Bahuguna, K. K. Gupta, K. Singh, "Expected number of intensity level crossings in a normal speckle pattern," J. Opt. Soc. Am. {\bf 70}, 874-876 (1980).

\bibitem{kurtz73}
 C. Kurtz, H. Hoadley, and J. DePalma, "Design and synthesis of random phase diffusers," J. Opt. Soc. Am. {\bf 63},1080-1092 (1973).

\bibitem{Goodman06}
 J. W. Goodman,{\em Speckle Phenomena in Optics: Theory and Applications} (Roberts, 2006).

\bibitem{Asakura}
 N. Takai, T. Iwai, T. Asakura, "Laser speckle velocimeter using a zero-crossing technique for spatially integrated intensity fluctuation," OPT. ENGNG. {\bf 20}, 320-325 (1981).
 
\bibitem{Asakura81}
T. Asakura, N. Takai, "Dynamic Laser Speckles and Their Application to Velocity Measurements of the Diffuse Object,"  Appl. Phys. {\bf 25}, 179-194 (1981).


\bibitem{Takai}
N.Takai, T. Iwai, T.Asakura, "Real-time velocity measurement for a diffuse object using zero-crossings of laser speckle," J. Opt. Soc. Am. {\bf 70}, 4  450-455 (1980).

\bibitem{Iwai}
T. Iwai, N. Takai, T. Asakura, "Simultaneous Magnitude and Direction Measurements of a Diffuse Object's Velocity Using the Rotating Directional Detecting Aperture in a Laser Speckle Zero-crossing Method,"  Optica Acta {\bf 28},  6  857-870  (1981).


\bibitem{Takai2}
N. Takai, Sutanto, T. Asakura, "Dynamic statistical properties of laser speckle due to longitudinal motion of a diffuse object under Gaussian beam illumination,"  J. Opt. Soc. Am. {\bf 70}, 7  827-834  (1980).

\bibitem{Barakat}
R. Barakat, "Zero-crossing rate of differentiated speckle intensity,"  J. Opt. Soc. Am. A {\bf 11}, 671-673 (1994).

\bibitem{Barakat88} 
R. Barakat, "Level-crossing statistics of aperture-integrated isotropic speckle," J. Opt. Soc. Am. A {\bf 5}, 1244-1247 (1988).

\bibitem{Yura10}
H. T. Yura and S. G. Hanson, "Mean level signal crossing rate for an arbitrary stochastic process," J. Opt. Soc. Am. A {\bf 27}, 4 797-807 (2010).


\bibitem{Barakat91}
R. Barakat, "Probability density of the radial gradient of aperture-averaged isotropic-speckle intensity," J. Opt. Soc. Am. A {\bf  8}, 2 450-451 (1991).

\bibitem{Ryden89}
 B. S. Ryden, A. L. Melott, D. A. Craig, J. R. Gott III, D. H. Wenberg, R. J. Scherrer, S. P. Bhavsar and J. M.  Miller, "The area of isodensity contours in cosmological models and galaxy surveys,"  The Astrophysical Journal {\bf 340}, 647 (1989).

\bibitem{ryden1988}
B. S. Ryden, "The area of isodensity contours as a measure of large-scale structure", The Astrophysical Journal {\bf333}, 41-44 (1988).

\bibitem{Matsubara96}
T. Matsubara, "Statistics of Isodensity Contours in Redshift Space,"  The Astrophysical Journal {\bf 457}, 13 (1996).
\bibitem{matsubara03} 
T. Matsubara, "Statistics of smoothed cosmic fields in perturbation theory,"  The Astrophysical Journal {\bf584}, 1-33  (2003).


\bibitem{sadegh11}
M. S. Movahed and S. Khosravi, "Level Crossing Analysis of Cosmic Microwave Background Radiation: A method for detecting cosmic strings," Journal of Cosmology and Astroparticle Physics {\bf1103}, 012 (2011). 




\bibitem{Beckmann67}
P. Beckmann, {\em Probability in Communication Engineering} (Hartcourt Brace \& World, 1967), Sec. 6.7.


\bibitem{Kakac1}
S. Kakac, A. Pramuanjaroenkij, " Review of convective heat transfer enhancement with nanofluids,"  Int. J. Heat and Mass Transf. {\bf 52}, 3187-3196 (2009).

\bibitem{Kakac}
 S. Kakac, A. Pramuanjaroenkij, "Single-phase and two-phase treatments of convective heat transfer
enhancement with nanofluids - A state-of-the-art review,"  International Journal of Thermal Sciences {\bf100}, 75-97 (2016).

\bibitem{Wong}
 E. Ebrahimnia-Bajestan, M. Charjouei Moghadam, H. Niazmand, W. Daungthongsuk, S. Wongwises, "Experimental and numerical investigation of nanofluids heat transfer characteristics for application in solar heat exchangers," International Journal of Heat and Mass Transfer {\bf92}, 1041-1052 (2016).

\bibitem{Duang}
W. Duangthongsuk, S. Wongwises, "Comparison of the effects of measured and computed thermophysical properties of nanofluids on heat transfer performance,"  Experimental Thermal and Fluid Science {\bf34}, 616-624 (2010).

\bibitem{Azmi}
 W. H. Azmi, K.V. Sharma, R. Mamat, G. Najafi, M. S. Mohamad, "The enhancement of effective thermal conductivity and effective dynamic viscosity of nanofluids: A review,"  Renewable and Sustainable Energy Reviews {\bf53}, 1046-1058 (2016).

\bibitem{Buongiorno}
 J. Buongiorno, D.C. Venerus, N. Prabhat, T. McKrell, J. Townsend, R. Christianson, "A benchmark study on the thermal conductivity of nanofluids,"  J. Appl. Phys. {\bf106}, 094312 (2009).

\bibitem{Wang}
X.-Q. Wang, A.S. Mujumdar, "Critical review of heat transfer characteristics nanofluids,"  Int. J. Therm. Sci. {\bf 46}, 1-19 (2007).


\bibitem{Guyot}
 S. Guyot, M. C. P\`{e}ron and E. Del\`{e}chelle, "Spatial speckle characterization by Brownian motion analysis," Phys. Rev.  E. {\bf70},  046618 1-8 (2004).

\bibitem{Chicea}
D. Chicea, "Speckle size, intensity and contrast measurement application in micron-size particle concentration assessment,"  The European Physical Journal Applied Physics {\bf 40}, 305-310 (2007).
\bibitem{aizu} Y. Aizu and T. Asakura, {\em Spatial Filtering Velocimetry: Fundamentals and Applications} , (Springer , 2006)


\bibitem{Scherrer91}
 R. J. Scherrer, E. Bertschinger, "Statistics of primordial density perturbations from discrete seed masses,"  The Astrophysical Journal {\bf 381}, 10 349-360 (1991).

\bibitem{Juszkiewicz95}
R. Juszkiewicz, D. H.  Weinberg, P. Amsterdamski, M. Chodorowski, F. Bouchet, "Weakly nonlinear Gaussian fluctuations and the edgeworth expansion," The Astrophysical Journal {\bf 442}, 39-56 (1995).




\bibitem{Bernardeau95}
 F. Bernardeau, L. Kofman, "Properties of the cosmological density distribution function," The Astrophysical Journal {\bf 443}, 479-498 (1995).
\bibitem{ma85}S.K. Ma, Statistical Mechanics (Philadelphia:World Scientific, 1985).



\bibitem{cas90} B. Castaing, Y. Gagne, and E. J. Hopfinger, "Velocity probability density functions of high Reynolds number turbulence," Physica D. {\bf 46}, 177 (1990).
\bibitem{chab94} B. Chabaud, A. Naert, J. Peinke, F. Chill\`{a}, B. Castaing, and B. H\'{e}bral, "A transition toward developed turbulence," Phys. Rev. Lett. {\bf73}, 3227 (1994).
\bibitem{aren98} A. Arneodo, E. Bacry, and J. F. Muzy,  "Random cascades on wavelet dyadic trees," J. Math. Phys. {\bf39}, 4142 (1998).
\bibitem{bac01} E. Bacry, J. Delour, and J. F. Muzy, "Multifractal random walk," Phys. Rev. E, {\bf64}, 026103 (2001).

\bibitem{Ghashghaie96} S. Ghashghaie, W. Breymann, J. Peinke, P. Talkner, and Y. Dodge, "Turbulent Cascades in Foreign Exchange Markets,"  Nature London {\bf381}, 767 (1996).
\bibitem{kiy06}K. Kiyono, Z. R. Struzik, and Y. Yamamoto, "Criticality and phase translation in stock-price fluctuation,"  Phys. Rev. Lett. {\bf96}, 068701 (2006).
\bibitem{kiy04} K. Kiyono, Z. R. Struzik, N. Aoyagi, S. Sakata, J. Hayano, and Y. Yamamoto, "Critical scale invariance in a healthy human heart rate," Phys. Rev. Lett. {\bf93}, 178103 (2004).
\bibitem{kiy05} K. Kiyono, Z. R. Struzik, N. Aoyagi, F. Togo, and Y. Yamamoto, "Phase translation in a healthy human heart rate,"  Phys. Rev. Lett. {\bf95}, 058101 (2005).
\bibitem{sol10} F. Shayeganfar,  S. Jabbari-Farouji,  M. Sadegh Movahed,  G. R. Jafari,  and M. Reza Rahimi Tabar, "Stochastic qualifier of gel and glass transitions in laponite suspensions," Phys. Rev. E, {\bf 81}, 061404 (2010).
\bibitem{sol13} F. Shayeganfar, M. Sadegh Movahed, G.R. Jafari, "Discrimination of Sol and Gel states in an aging clay suspension," Chemical Physics {\bf 423}, 167-172 (2013).

\bibitem{kohi15} Z. Koohi Lai, S. Vasheghani Farahani, S. M. S. Movahed, G. R. Jafari, "Coupled uncertainty provided by a multifractal random walker," Physics Letters A 379, 2284-2290 (2015).

\bibitem{Karimzadeh13}
R. Karimzadeh, M. Arshadi, "Thermal lens measurement of the nonlinear phase shift and convection velocity,"  Lasar Phys. {\bf 23}, 115402 (2013).

\bibitem{gold99} W. I. Goldburg, "Dynamic light scattering," American Journal of Physics, {\bf 67}, 1152 (1999).

\bibitem{moun66} R. D. Mountain, "Spectral Distribution of Scattered Light in a Simple Fluid," Rev. Mod. Phys. {\bf 38}, 205-214 (1966).
\bibitem{scha72} D. W. Schaefer and B. J. Berne, "Light Scattering from Non-Gaussian Concentration Fluctuations,"  Phys. Rev. Lett. {\bf 28}, 475-478 (1972).
\bibitem{bern74} B. J. Berne and R. Pecora, "Laser Light Scattering from Liquids," Annu. Rev. Phys. Chem. {\bf 25}, 233-253 (1974).
\bibitem{pathria11} R.K. Pathria and Paul D. Beale, {\em Statistical Mechanics}, Third Edition, 2011.

\bibitem{Pratt73}
 A. R. Pratt, {\em Some theoretical considerations concerning time statistics in signal detection}, in Signal Processing, Edited by J. Griffiths, P. Stocklin, and C. van Schooneveld, (Academic, New York, 1973).


\bibitem{brar11} S. K. Brar and M. Verma, "Measurement of nanoparticles by light scattering techniques," Trends Anal. Chem. {\bf 30}(1), 4-17 (2011).

\bibitem{braun11} A. Braun, O. Couteau, K. Franks, V. Kestens, G. Roebben, A. Lamberty and T. P. J.  Linsinger, "Validation of dynamic light scattering and centrifugal liquid sedimentation methods for nanoparticle characterisation,"  Advanced Powder Technology, {\bf 22}(6), 766-770 (2011). 
 
 \bibitem{chic08} D. Chicea, "Nanoparticle sizing by coherent light scattering computer simulation results,"  Journal of Optoelectronics and Advanced Materials, {\bf 10}, 4 (2008).
 

\bibitem{boyd11} R. D. Boyd, S. K. Pichaimuthu and A. Cuenat, 
"New approach to inter-technique comparisons for nanoparticle size measurements: using atomic force microscopy, nanoparticle tracking analysis and dynamic light scattering," Colloids Surf., A Physicochem. Eng. Aspects, {\bf 387} (1-3), 35-42 (2011). 






\end{thebibliography}
\end{document}